\documentclass[sigconf, nonacm]{acmart}

\setcopyright{none} 

\usepackage{times}
\usepackage{helvet}
\usepackage{courier}
\usepackage{graphicx}
\usepackage{makecell}
\usepackage{multirow}
\graphicspath{{Images/}}
\usepackage{booktabs} 
\usepackage[ruled]{algorithm2e} 
\usepackage{tabularx}

\SetAlFnt{\small}
\SetAlCapFnt{\small}
\SetAlCapNameFnt{\small}
\SetAlCapHSkip{0pt}
\IncMargin{-\parindent}

\usepackage{makecell}
\usepackage{lscape} 
\usepackage{threeparttable}  
\usepackage{booktabs}
\usepackage{multirow}

\setcitestyle{authoryear}

\copyrightyear{2022}
\acmYear{2022}
\setcopyright{rightsretained}
\acmConference[AIES'22]{Proceedings of the 2022 AAAI/ACM Conference
on AI, Ethics, and Society}{August 1--3, 2022}{Oxford, United Kingdom}
\acmBooktitle{Proceedings of the 2022 AAAI/ACM Conference on AI,
Ethics, and Society (AIES'22), August 1--3, 2022, Oxford, United
Kingdom}\acmDOI{10.1145/3514094.3534202}
\acmISBN{978-1-4503-9247-1/22/08}

\begin{document}
\title[What are People Talking about in \#BlackLivesMatter and \#StopAsianHate?]{What are People Talking about in \#BlackLivesMatter and \#StopAsianHate? Exploring and Categorizing Twitter Topics Emerging in Online Social Movements through the Latent Dirichlet Allocation Model}
\subtitle{Accepted at AAAI/ACM Conference on AI, Ethics, and Society (AIES’22), August 1–
3, 2022, Oxford, United Kingdom.}

\author[Xin Tong*]{Xin Tong}
\affiliation{%
  \department{Art and Humanity Division}
  \institution{Duke Kunshan University}
  \country{China}
}
\authornotemark[1]
\authornotemark[2]

\author{Yixuan Li}
\affiliation{%
  \institution{Duke Kunshan University}
  \country{China}
}

\authornotemark[3]

\author{Jiayi Li}
\affiliation{%
  \institution{Duke Kunshan University}
  \country{China}
}

\author{Rongqi Bei}
\affiliation{%
  \institution{Duke Kunshan University}
  \country{China}
}

\author[Luyao Zhang*]{Luyao Zhang}
\affiliation{%
  \department{Data Science Research Center and Social Science Division}
  \institution{Duke Kunshan University}
  \country{China}
}

\authornote{Corresponding authors: \newline
 Luyao Zhang (email: lz183@duke.edu, institutions: Data Science Research Center and Social Science Division, Duke Kunshan University) and  Xin Tong (email: xt43@duke.edu, institutions: Art and Humanity Division, Duke Kunshan University)}
\authornote{These authors contributed equally to this work.}
\authornote{Also with SciEcon CIC, 71-75 Shelton Street, Covent Garden, London, United Kingdom, WC2H 9JQ}

\begin{abstract}
Minority groups have been using social media to organize social movements that create profound social impacts. Black Lives Matter (BLM) and Stop Asian Hate (SAH) are two successful social movements that have spread on Twitter that promote protests and activities against racism and increase the public's awareness of other social challenges that minority groups face. However, previous studies have mostly conducted qualitative analyses of tweets or interviews with users, which may not comprehensively and validly represent all tweets. Very few studies have explored the Twitter topics within BLM and SAH dialogs in a rigorous, quantified and data-centered approach. Therefore, in this research, we adopted a mixed-methods approach to comprehensively analyze BLM and SAH Twitter topics. We implemented (1) the latent Dirichlet allocation model to understand the top high-level words and topics and (2) open-coding analysis to identify specific themes across the tweets. We collected more than one million tweets with the \#blacklivesmatter and \#stopasianhate hashtags and compared their topics. Our findings revealed that the tweets discussed a variety of influential topics in depth, and social justice, social movements, and emotional sentiments were common topics in both movements, though with unique subtopics for each movement. Our study contributes to the topic analysis of social movements on social media platforms in particular and the literature on the interplay of AI, ethics, and society in general.

\end{abstract}


\maketitle
\begin{figure}[htp]
  \centering
  \includegraphics[width=8cm]{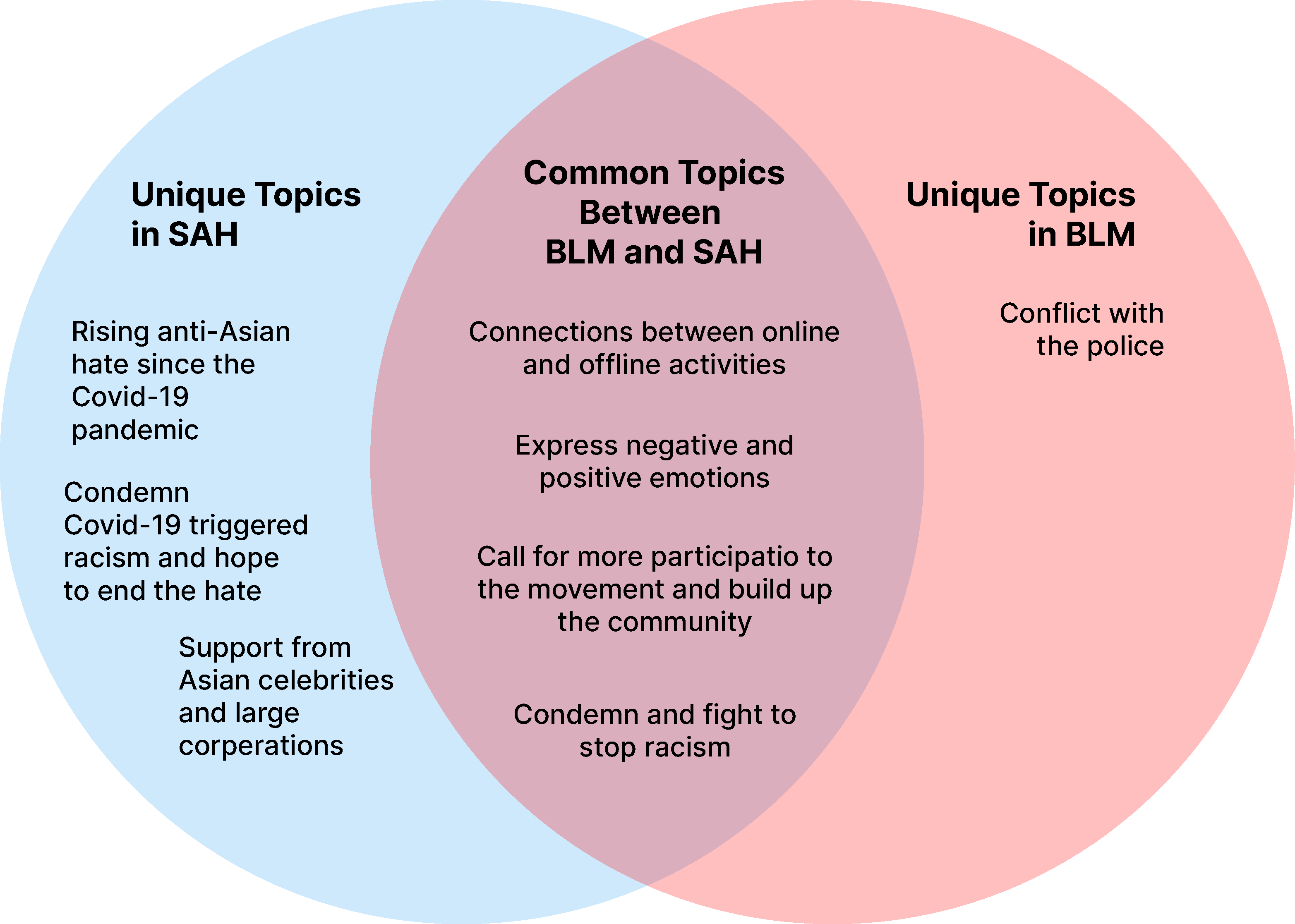}
  \caption{Common topics between BLM and SAH, and unique topics in the two movements}
  \label{fig:9}
\end{figure}
\section{Introduction}
Minority groups, vulnerable populations, and those advocating for them (e.g., Black and Asian communities in the U.S., members of the LGBTQ+ community, and feminists) ~\cite{klassen_more_2021,lucero_safe_2017,mueller_demographic_2020,harley_online_2014} have been using social computing technologies and platforms to organize their online and offline social movements, mobilize supporters, express political appeals, reframe issues and create profound social impacts.
\par
Black Lives Matter (BLM) is one of the many successful social movements that has started and expanded on social media ~\cite{mundt_scaling_2018}. The movement started in 2013 ~\cite{dacon_what_2021}. BLM aims to bring attention to the racism and inequality that Black people face and protest against police brutality and other forms of violence motivated by racial discrimination ~\cite{taylor_blacklivesmatter_2016}. The BLM movement has recently experienced an enormous surge in public attention due to the murder of George Floyd and the ensuing protests~\cite{barrie_searching_2020}. The hashtag \#blacklivesmatter was used roughly 47.8 million times on Twitter in the two weeks following the tragedy~\cite{anderson_blacklivesmatter_2020}, and polls estimated that between 15 million to 26 million people participated in various protests related to George Floyd’s murder ~\cite{buchanan_black_2020}.
\par
Approximately one year later, the shooting death of six Asian women in Atlanta in March 2021 led to another online social movement. The Atlanta shooting was believed to be a racially motivated hate crime and one of many incidents of violence against Asians since the outbreak of the COVID-19 pandemic. Similar to the impact of George Floyd’s murder on the BLM movement, these murders triggered the Stop Asian Hate (SAH) social movement on Twitter using the \#stopasianhate hashtag ~\cite{lee_questing_2021}. Since the pandemic began, ~\cite{bernardino_center_2021}, hate crimes against Asians have increased 150\%, and Stop AAPI Hate~\cite{yam_anti-asian_2021} has received 3,795 discriminatory incident reports. The SAH movement united several rallies against anti-Asian violence in response to racism toward Asian Americans~\cite{redd_unity_2021}.
\par
Prior studies~\cite{peng_event-driven_2019,mundt_scaling_2018,ince_social_2017} have collected different data for topic analysis and utilized varied analysis methods to understand the massive number of tweets with the \#blacklivesmatter hashtag. Some researchers ~\cite{ince_social_2017} have looked into how users have engaged with \#blacklivesmatter and/or \#stopasianhate Twitter topics, and only a few~\cite{dacon_what_2021,gallagher_divergent_2018} have investigated the topics being discussed on Twitter. For instance, Dacon and Tang conducted thematic text analysis and studied users’ tweet topics and associated attitudes. However, they included only tweets before July 2020 and lacked a data-centered approach to comprehensively analyze a large number of tweets with the BLM hashtag~\cite{dacon_what_2021}.. In another example, although Gallagher et al. performed word- and topic-level data analysis, the authors treated the \#blacklivesmatter hashtag as a prompt/proxy but did not analyze the content of the tweets~\cite{gallagher_divergent_2018}. Similarly, to our knowledge, only a few studies have examined the topics that emerged in tweets with the \#stopasianhate hashtag~\cite{lee_questing_2021}.
\par
Moreover, online social movements have been proven to have a profound impact on offline protests~\cite{peng_event-driven_2019}, but little work has been done to understand the potential relationships between BLM/SAH’s online (Twitter) activities and offline social events/protests and the impacts one had over each other. Therefore, we aim to adopt a data-centered mixed-methods approach to explore the range of topics in tweets with \#blacklivesmatter and \#stopasianhate and further analyze the potential relationship between the real-life social movements and online BLM/SAH activities.
\par
Both the \#blacklivesmatter and \#stopasianhate Twitter movements were triggered or reinitiated due to racism-related murders, and both attracted more attention worldwide in the past two years than they had previously. However, to our knowledge, no studies have compared the similarities and differences of tweet topics in these two movements. SAH started 8 years later than BLM and comparatively speaking, lacks a strong online community. Therefore, we want to compare these two social movements on Twitter and provide insights into how the characteristics of the minority communities involved might affect the online and offline protests and other social activities.
\par
More specifically, our research questions (RQs) are as follows: \\
(1)	What are the main topics being discussed in tweets with \#blacklivesmatter and \#stopasianhate hashtags? \\
(2)	What are the connections between the \#blacklivesmatter and \#stopasianhate social movements on Twitter and the related offline social movements/events? \\
(3) What are the differences and similarities in the topics that emerged in the online movements using these two hashtags? \\
To answer these RQs, we collected 1,263,683 tweets with the \#blacklivesmatter hashtag from May to December 2020 and 96,691 tweets with the \#stopasianhate hashtag from March to December 2021. We adopted a mixed-methods approach, including (1) the data-centered latent Dirichlet allocation (LDA) model~\cite{blei_latent_2003,chuang_termite_2012,sievert_ldavis_2014} to analyze the topics from both online social movements and (2) a qualitative open-coding approach for approximately 1,700 tweets selected proportionally to our BLM and SAH datasets.
\par
Findings our study constructed a comprehensive picture of the major topics in the \#blacklivesmatter and \#stopasianhate movements on Twitter from a massive corpus of tweets. Both movements include tweets denouncing racism, calling for participation in the respective movements, connecting online and offline events, and expressing emotional responses. Additionally, the BLM and SAH movements include unique subtopic(s). \#blacklivesmatter concerns the public’s discontent with police brutality, and \#stopasianhate touches on the negative effects on the Asian community due to the COVID-19 pandemic and received support from various pop stars. Our results also demonstrated that offline social movement activity usually occurred concurrently with the volume peaks for \#blacklivesmatter and \#stopasianhate tweets according to their timestamps. Moreover, by examining these two recent and important online social movements, \#blacklivesmatter and \#stopasianhate, our study provides insights into the similarities and differences of tweet topics in both social movements and clarifies the roles of social movements’ online activities and offline protests. Our research makes two major contributions. First, our research examines the links between online and offline movements, as well as the content of Twitter topics, which sheds light on the relevance of social media in uniting people who share common concerns. Second, we use quantitative (i.e., the LDA model, network) and qualitative (i.e., open coding) methodologies to identify important themes in the BLM and SAH online social movements that span the major issues of debate on Twitter.
\section{Related Work}
\subsection{Black Lives Matter Movement and Stop Asian Hate Movement on Social Media}
The inception and increasing popularity of the Black Lives Matter (BLM) online movement were both enabled by the existence of so-called Black Twitter, an online community hosted on Twitter that embraces and celebrates content oriented around Black culture~\cite{klassen_more_2021}. Even though most Black Twitter users are identified as African Americans, nonblack people are occasionally active in the online community, suggesting a solid understanding of Black culture among a diverse audience~\cite{brock_blackhand_2012}. According to ~\citeauthor{brock_blackhand_2012}~\shortcite{brock_blackhand_2012}, members of Black Twitter gather online to create and distribute information about Black culture, discuss issues relevant to the Black community, cultivate reciprocity within the community, and reflect on the shared experience of Black people. Whereas 21 percent of white adults and 25 of Hispanic adults use Twitter, over 24 percent of all African American adults use Twitter~\cite{perrin_share_2019}, Black people on Twitter formed an extremely connected and influential network by creating tight clusters~\cite{manjoo_how_2010}, which later led to the community’s swift response to several important events within the BLM movement.
\par
Activities on Black Twitter have centered around “Blacktags”~\cite{brock_blackhand_2012}, a series of race-related hashtags such as \#blacklivesmatter, \#ifsantawasblack, and \#onlyintheghetto~\cite{sharma_black_2013}. These Blacktags help curate content related to a particular topic~\cite{messina_groups_2007}], encourage the viral circulation of tweets and retweets~\cite{sharma_black_2013}, and further enable public/private conversations and larger social movements on Black Twitter ~\cite{clark_tweet_2014}. Previous studies~\cite{graham_content_2016,lee_black_2017} have revealed Black Twitter’s role as a counterpublic that reacts to the biased perspectives on Black people in the mainstream media, establishing an avenue through which Black people can challenge the bias they encounter and reconstruct and control their representation in cyberspace.
\par
In comparison to the BLM movement, the Stop Asian Hate (SAH) movement has been around for a significantly shorter period of time~\cite{lyu_state-level_2021} and lacks the backing of a strong online community such as Black Twitter. Due to the newness of the SAH movement, it has yet to obtain a significant place in academic discourse, with only a few studies~\cite{lyu_state-level_2021,lee_questing_2021,allen_pictures_2021} looking into how Twitter users have engaged in the SAH movement digitally. ~\citeauthor{lee_questing_2021}’s~\shortcite{lee_questing_2021} work was the only one we found that studied the content of \#stopasianhate conversations on Twitter. They used structural topic modeling (STM) to analyze 259,456 English tweets with the \#stopasianhate hashtag to understand the movement’s most salient topics (e.g., community building for anti-Asian violence, stopping violence and calling for support) on a daily basis, but their analysis covered only a seven-day period~\cite{lee_questing_2021}.
\par
Prior studies ~\cite{peng_event-driven_2019,mundt_scaling_2018,ince_social_2017,lyu_state-level_2021,lee_questing_2021,allen_pictures_2021} have investigated Twitter’s influence on the \#blacklivesmatter and \#stopasianhate online movement and offline protests. Although these studies~\cite{carney_all_2016,klassen_more_2021,lyu_state-level_2021} have demonstrated the ways in which Black people have played an active role in the \#blacklivesmatter and \#stopasianhate online movements, most studies have adopted qualitative approaches and focused on the participants of these social movements and their roles in the movement network. A few works have adopted a data-centered approach and mainly focused on users’ roles, interactions, and behaviors in movements~\cite{ince_social_2017,twyman_black_2017}. For instance, \citeauthor{ince_social_2017} \shortcite{ince_social_2017} analyzed a total of 66,159 tweets with the \#blacklivesmatter hashtag, and their findings suggested that Twitter users maintain decentralized interactions—a distributed framing—when participating in the BLM movement. In another work, \citeauthor{twyman_black_2017} \shortcite{twyman_black_2017} focused on users’ collective behaviors in documenting historical and contemporary events on Wikipedia and found that the participants’ collective actions supported the coverage of new events and the reexamination of preexisting knowledge. In \citeauthor{lyu_state-level_2021}'s \shortcite{lyu_state-level_2021} work, a study of public opinion on the \#stopasianhate and \#stopaapihate movement was conducted based on 46,058 Twitter users. Results showed that the movements mostly attracted women, younger adults, and members of the Asian and Black communities.
\subsection{Microblogging Social Media Data Analysis}
Microblogging is a relatively new form of communication that allows users to share brief text updates (often fewer than 200 characters) with various recipients through text messaging, instant messaging (IM), email, or the web ~\cite{java_why_2009}, e.g., tweets on Twitter or posts on Facebook. The minimal effort and time required to create a microblog encourage users to frequently share news and events ~\cite{muralidharan_hope_2011}, enabling them to rapidly communicate instantaneous messages to specific groups or global audiences~\cite{williams_what_2013}. Hashtags, denoted by a hash symbol ("\#"), are used as information organizers on varied microblogging platforms, where they help connect members of online communities by linking content with similar topics, allowing users to search within topics of interest, and offering overviews of trending topics ~\cite{yang_we_2012}.
\par
Twitter data have been widely used in research studies in various areas, including but not limited to politics and government, health care, education, business and finance, journalism, and eyewitness accounts of news stories~\cite{dann_twitter_2010,rathore_social_2017}. ~\citeauthor{williams_what_2013}~\shortcite{williams_what_2013} identified four aspects of Twitter data that researchers can consider: \emph{message} (the content that users enter and the associated metadata),
\emph{user} (the user’s digital identities exposed on Twitter), \emph{technology} (the underlying hardware implementations, APIs, and the software design and development) and \emph{concept} (encompassing introductory overviews and other discussion pieces). In our study, we focused on the message aspect of the Twitter data gathered under the \#blacklivesmatter hashtag, as few prior works have focused on the content of conversations related to the movement. Our work aims to provide deeper insights into how Twitter users participate in online social movements by evaluating the associated metadata (e.g., the date and time each tweet was posted) and performing topic modeling on the message content.
\par
A variety of machine learning (ML) and deep learning (DL) models, such as the discriminative multinomial naïve Bayes model (DMNBText) ~\cite{gokulakrishnan_opinion_2012}, support vector machine model (SVM) and its variants~\cite{coletta_combining_2014}, long short-term memory model (LSTM), recurrent neural network model (RNN), and text convolutional neural network model (TextCNN)~\cite{sequeira_large-scale_2019}, have been adopted to perform Twitter and other social media data analysis. ~\citeauthor{sequeira_large-scale_2019}~\shortcite{sequeira_large-scale_2019} experimented with various ML/DL models (e.g., naive Bayes, SVM, random forest, logistic regression, LSTM, RNN, RCNN, TextCNN, etc.) and performed text classification on the tweets of approximately 0.42 million users referencing drug abuse (DA) to characterize the spread of prescription DA-related tweets. ~\citeauthor{adrover_identifying_2015}~\shortcite{adrover_identifying_2015} performed sentiment analysis with boosted decision trees with AdaBoost, SVM, boosted decision trees with bagging, and other artificial neural networks to identify the adverse effects of HIV drug treatment.
\par
To examine the content discussed in the BLM online movement and SAH online movement, topic modeling, a sort of model used to uncover abstract subjects within a collection of text documents, is considered a viable strategy for providing interpretable summaries of the most intensely debated topics on Twitter. A variety of methods can be found in prior social media analysis, such as the topic leaders and coopetition model~\cite{sun_evoriver_2014}, latent semantic indexing (LSI)~\cite{huang_scalable_2014}, multilevel text analysis~\cite{dacon_what_2021} and latent Dirichlet allocation (LDA) ~\cite{wang_exploring_2014}. Despite the fact that both LSI and LDA have accessible API libraries, a previous study ~\cite{cvitanic_lda_2016} revealed that LDA produces more accurate results on larger datasets than LSI. LDA was first proposed by ~\citeauthor{blei_latent_2003}~\shortcite{blei_latent_2003} as a generative probabilistic model for evaluating topic probability as an explicit representation of text corpora. It was later integrated into an assortment of visualization techniques ~\cite{chuang_termite_2012,sievert_ldavis_2014} that help identify abstract topics in a given document. In our study, LDA was first adopted to identify the most hotly debated topics in the BLM and SAH online movements.
\par
A few studies investigated the topic being discussed in the tweets with \#blacklivesmatter hashtag~\cite{dacon_what_2021,gallagher_divergent_2018}. ~\citeauthor{dacon_what_2021} et al. identified the most central topics (e.g. police, protests, supports, and stories) and the topics' affiliated attitudes towards the protest and counter protests within the \#blacklivesmatter movement through thematic analysis~\cite{nowell_thematic_2017}. However, their investigation only covered the data before July 2020, and their work lacked testing metrics that could validate their model's performance~\cite{dacon_what_2021}. ~\citeauthor{gallagher_divergent_2018}~\shortcite{gallagher_divergent_2018} performed word-level analysis and topic-level analysis on 860,000 tweets collected from a one-year period from 2014 to 2015 to demonstrate the divergent discourse between protests and counter-protests under the \#blacklivesmatter hashtag and \#alllivesmatter hashtag. In their study, hashtags were regarded as proxies for topics; however, little evidence supported the notion that hashtag could accurately represent the abstract content of tweets~\cite{gallagher_divergent_2018}. Moreover, to our knowledge, there is an absence of comprehensive content-wise analysis of tweets with \#stopasianhate hashtag. Given the extraordinary level of participation in both \#blacklivesmatter and \#stopasianhate online social movements and offline social events locally and globally~\cite{giorgi_twitter_2020}, we would like to investigate the topics that have been discussed in the tweets with \#blacklivesmatter and \#stopasianhate hashtags and further compare the differences and similarities in these two recent racial-related social movements. 
\section{Method}
We release data and code as open access on the Github repository: \url{https://github.com/HCI-Blockchain/Blacklivesmatter}.
\subsection{Data}
Here, we introduce the techniques and criteria we used to collect our datasets for the two online social movements and provide details of these two datasets.
\par
Previous works selected different date ranges, data sizes, and sampling methods as their data selection criteria. For example,  ~\citeauthor{dacon_what_2021}~\shortcite{dacon_what_2021} randomly sampled 12,500 English tweets from 37 million raw tweets with the \#blacklivesmatter, \#alllivesmatter, or \#bluelivesmatter hashtags from 2013 to 2020. Gallagher et al. ~\cite{gallagher_divergent_2018} randomly selected 10\% of English tweets from a total of 767,139 tweets with the \#blacklivesmatter or \#alllivesmatter hashtags from 2014 to 2015. In another example, Carney et al. ~\cite{carney_all_2016} selected 100 samples out of the initial 500 English tweets on murders of Michael Brown and Eric Garner from December 2nd to 7th,
2014.  ~\citeauthor{lee_questing_2021}~\shortcite{lee_questing_2021} selected 259,456 English tweets on anti-
Asian hate, anti-Asian crimes, and support for Asian communities from March 16, 2021, to March 22, 2021. In summary, previous studies have used diverse date ranges for data selections, ranging from several days to years, and different raw data sizes, ranging from several hundreds to millions. However, with large datasets, researchers \cite{dacon_what_2021,carney_all_2016} tend to apply random sampling methods to lower the data size.
\par
Below are the selection criteria for our paper:  
Although Twitter API \footnote{https://developer.twitter.com/en/docs/twitter-api} has been widely adopted to crawl Twitter data previously~\cite{lee_questing_2021,panda_covid_2020}, this API has an upper limit of 180 queries per 15 minutes. Therefore, in our research, we decided to use the Snscrape API\footnote{https://github.com/JustAnotherArchivist/snscrape} to collect tweets related to \#blacklivesmatter, and \#stopasianhate because it outperforms Twitter API in terms of the comprehensiveness of its queries.
\par
We queried two datasets, one for tweets with \#blacklivesmatter (Dataset B) and the other for tweets with \#stopasianhate (Dataset S). We excluded all retweets and only included tweets that satisfied the following inclusion criteria:
(1)	The tweet is English. Considering the computational cost of including all the tweets, we choose "corona lang" as "en" (English) in the snscrape API for querying English tweets.
(2)	For Dataset B, the tweet has the \#blacklivesmatter hashtag, and for Dataset S, the tweet has \#stopasianhate hashtag.
(3)	The tweet received at least three likes. Positive social feedback (likes and shares) accentuates the inclination of online moral outrage expressions ~\cite{brady_how_2021}. Therefore, applying criteria (3) helps select tweets that convey more common viewpoints and compiling a more representative dataset.
\par
For Dataset B, we collected tweets with the \#blacklivesmatter hashtag starting from May 25, 2020, the day of George Floyd’s death, which prompted a surge of protests related to this online social movement. The end date for tweets in Dataset B is December 31, 2020. Similarly, for Dataset S, we collected tweets with the \#stopasianhate hashtag between March 16, 2021, the day after the Atlanta spa shooting incident, and December 31, 2021. Dataset B consists of 1,263,704 tweets in total, and Dataset S includes a much smaller number of tweets, 96,958 in total. 
\par
\begin{figure}[htp]
    \centering
    \includegraphics[width=6cm]{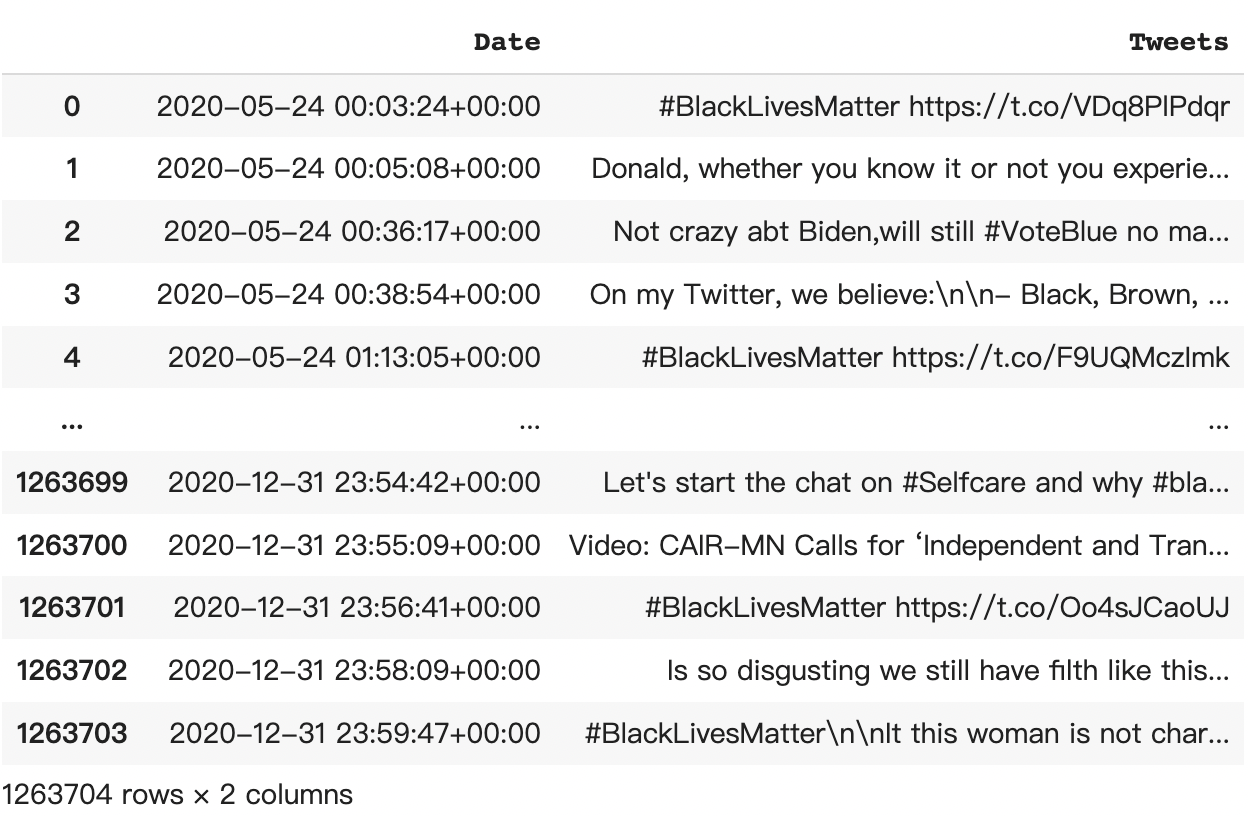}
    \caption {Sample Dataset}
    \label{fig:6}
\end{figure}

In the Appendix, Table ~\ref{table:1} presents the data dictionary for both datasets, and Figure ~\ref{fig:6} shows a few sample items from our final datasets. Both datasets contain the tweet ID, tweet date (with daily frequency), and tweet content. We filtered out retweets and included only original tweets to better analyze Twitter users’ unique topics and reduce the potential for amassing repetitive topics. For data preprocessing, we first tokenized the tokens and removed stopwords and punctuation. In addition, we applied lemmatization to the raw words and generated a document-term pair for LDA model input.
\subsection{Top Words and Network Analysis Methods}
Previous studies have used quantitative network analysis for the online topic modeling of social movements. For instance, ~\citeauthor{jackson_hashtagactivism_2020}~\shortcite{jackson_hashtagactivism_2020} analyzed social media activism by calculating the network size, centrality, and temporal relationships among different hashtags. They used networks to visualize how individual Twitter accounts are connected by retweets and mentions and then qualitatively analyzed the content of tweets. Other researchers have applied the theory of network-level reciprocal disclosures to the hashtag activism of the \#MeToo movement \cite{gallagher_divergent_2018}, which concerns sexual violence. They detailed how survivors disclosed diverse sexual violence experiences and interacted with others within the \#MeToo network. Another study by Dacon and Tang~\cite{dacon_what_2021} first applied word clouds to interpret the significance of textual information and ranked the top 200 most frequently used bigrams. They also applied multilevel text analysis to the raw qualitative tweets and utilized the Python package NetworkX7 to manipulate their tweet network structures, calculating the degree centrality and betweenness centrality of the network.
\par
In this study, we refer to the methodology used by Dacon and Tang \cite{dacon_what_2021}. We first plot the top most frequently used words in the BLM and SAH datasets and create word clouds to provide the readers with a general understanding of the online conversations within these movements. Different from Dacon and Tang’s research on visualizing thematic analysis results through networks, we use the network to visualize the word bigram ranking. We followed procedures mentioned by ~\citeauthor{morrissey_analyze_2018}~\shortcite{morrissey_analyze_2018} and paired up words in cleaned tweets according to their co-occurrence across the tweets. We utilized networks to visualize how keywords are linked to make an absolute comparison between word-word pairs. A counter is defined to capture the top 50 most frequent co-occurrences. We used the Python package NetworkX\footnote{https://networkx.org/} for visualizing weighted graphs for \#blacklivesmatter and \#stopasianhate tweets separately. In the graphs, each node represents a word, each edge represents a connection between co-occurring words, and the thickness of edges depicts the absolute frequency of word pair. To position the nodes in the graph, we implemented NetworkX’s "spring layout" function, a method based on the Fruchterman–Reingold force-directed algorithm ~\cite{fruchterman_graph_1991}. 
\subsection{Latent Dirichlet Allocation Topic Analysis Method}
To identify the topics for both online social movements, we utilized latent Dirichlet allocation (LDA) ~\cite{blei_latent_2003,chuang_termite_2012,sievert_ldavis_2014}, a probabilistic model that can be used to form a high-level understanding of the content topics. The LDA model is an unsupervised machine learning model that discovers a mixture of “topics” by analyzing the semantic structure of a document and returning a probability distribution over words that characterize each topic. Each topic contains a set of keywords defining the topic, and the text tokens are distributed over latent topics among the documents. We can interpret the estimated parameters as clusters of words that are likely to occur together in messages.
\par
The LDA model has three hyperparameters $\alpha$ and $\beta$ and the number of topics (k). Here, $\alpha$ controls per-document topic distribution, and $\beta$ determines per topic word distribution. More intuitively, the higher the $\alpha$, the more topics created per document; The higher the $\beta$, the more keywords created per topic. Therefore, given parameters $\alpha$ and $\beta$, the joint distribution of a topic mixture $\theta$, a set of z topics , and a set of N words. The probability of a corpus is calculated as:
\begin{figure}[htp]
    \centering
    \includegraphics[width=6.5cm]{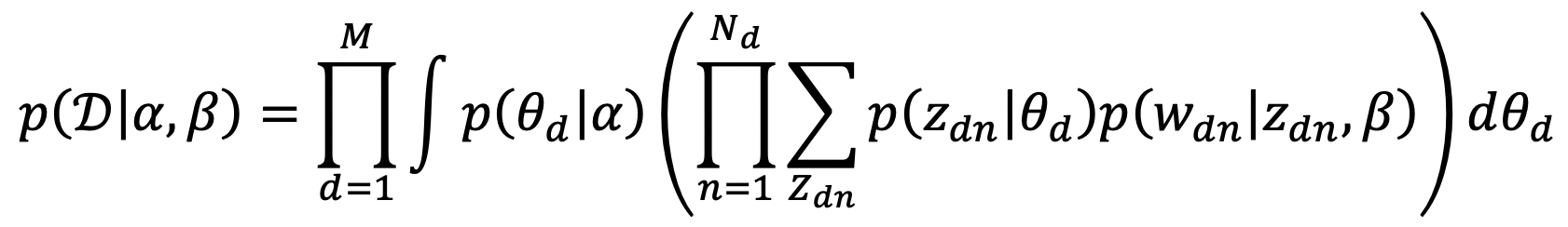}
    \label{fig:2}
\end{figure}
The training corpus generated by tokenized Tweets is used as data input. Models are trained with different topic numbers, $\alpha$, and $\beta$ parameters. We use a coherence score to evaluate the LDA model and pick the hyperparameters that give the best performance. Intuitively, the coherence score can evaluate whether the topics discovered are semantically interpretable or just artifacts of statistical inferences. The higher the coherence score, the more interpretable the model is. Technically, the measure of the coherence score depends on a sliding window, a one-set segmentation of the top words, and an indirect confirmation measure that uses normalized pointwise mutual information (NPMI) and the cosine similarity ~\cite{sun_conversational_2020}. 
\par
\begin{figure}[htp]
    \centering
    \includegraphics[width=9cm]{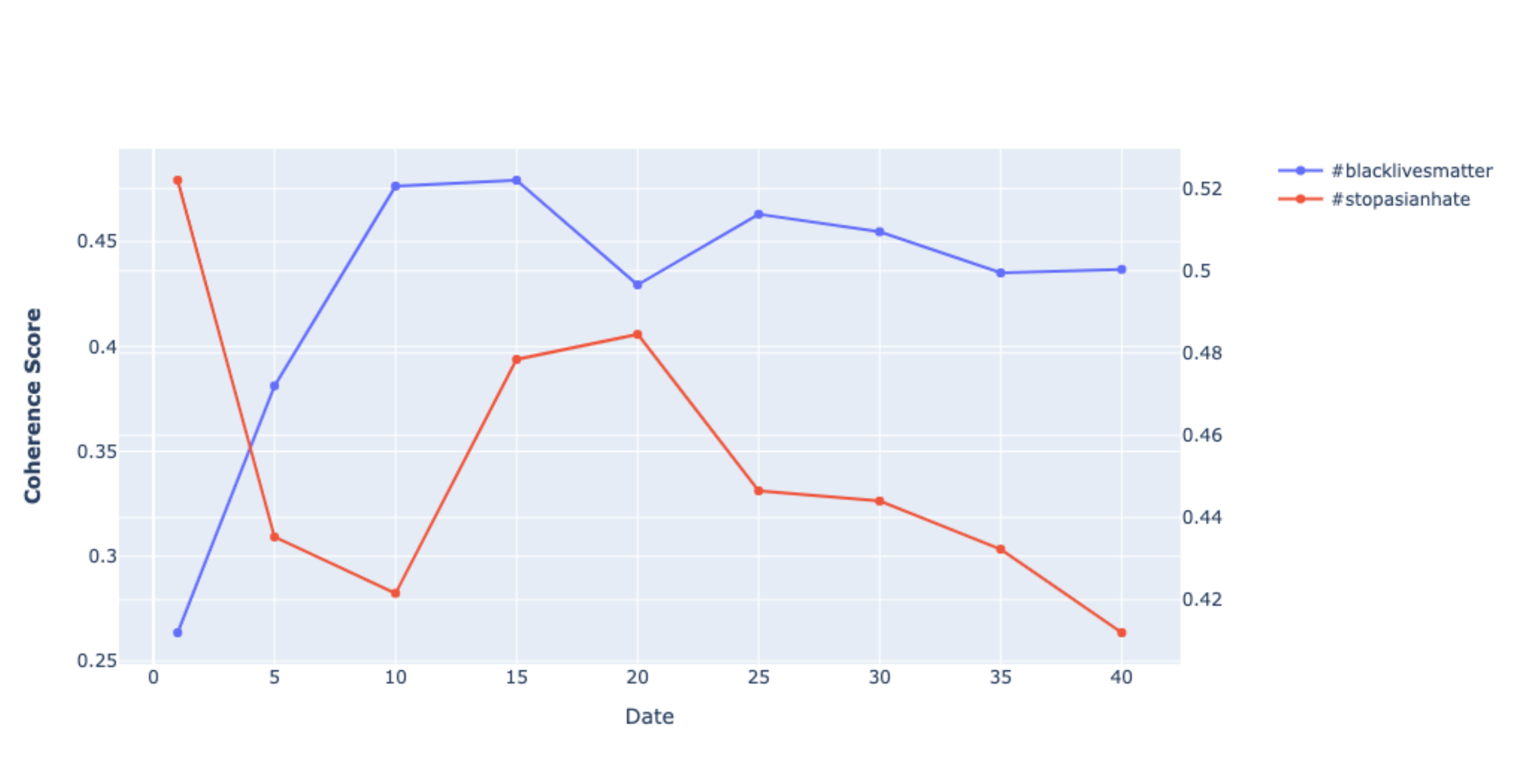}
    \caption{Coherence Score of Different Number of Topics for \#blacklivesmatter and \#stopasianhate}
    \label{fig:9.1}
\end{figure}

While adopting the LDA model for topic analysis, we varied the targeted number of topics being explored and calculated the coherence scores for the following targeted number of topics: 5, 10, 15, 20, 25, 30, 35, 40. We observed that for the \#blacklivesmatter dataset, the highest coherence score is k=15. According to Figure ~\ref{fig:9.1}, for \#stopasianhate, the highest coherence score is achieved at k=20. Therefore, we select the model with the highest coherence score: 15 topics for \#blacklivesmatter and 20 topics for \#stopasianhate.
\par
After determining the topic number, we fine-tuned hyperparameters $\alpha$ and then further calculated the coherence scores by following the procedure proposed by Pathik etal. ~\cite{blei_latent_2003}. We consider values of $\alpha$ and $\beta$ ranging from 0.01 to 0.91, with 0.30 as regular intervals. After we ran different combinations of $\alpha$ and $\beta$,we compiled the coherence scores in Table ~\ref{tab:2}). Finally, we chose the best-fitting hyperparameters by comparing the coherence score. For the \#blacklivesmatter dataset, the pair, $\alpha$=0.31 and $\beta$=0.91 reached the highest coherence score of 0.512 for 15 topics. For the \#stopasianhate dataset, the same pair $\alpha$ and $\beta$ reached the highest coherence score of 0.483 for 20 topics. Once the topics were identified by the LDA model, human coders applied open coding methods to discover the themes from the LDA results.
\subsection{Qualitative Thematic Analysis}
We conducted qualitative thematic analysis based on the LDA results to gain a more comprehensive understanding of the topics addressed in the BLM and SAH movements. To compile tweets for analysis, we randomly sampled 5 tweets for each keyword obtained under the list of obtained topics with the optimal coherence scores from the two movements. The compilation yielded 600 tweets for the BLM movement and 1175 tweets for the SAH movement. Then, two coders (two of the authors) independently performed inductive open coding \cite{corbin_grounded_1990} on tweets in the first topic from the sample tweet sets of BLM and SAH. Then, the two coders met to compare their independent coding results and discuss the codes. If the two coders failed to reach a consensus, a third researcher joined the discussion to help make a final decision. Eventually, researchers merged the mutually agreed code to create a codebook. They then populated the codebook with the remainder of the sampled tweets. Finally, they used affinity diagramming~\cite{hartson_ux_2012} to group comparable codes and determine common themes between the two movements as well as unique themes in each. For the BLM movement, we developed 42 codes; for the SAH movement, we developed 87 codes. We also identified 4 shared themes in both movements, as well as 1 unique theme in the BLM movement and 3 unique themes in the SHA movement (see Figure~\ref{fig:9} for themes).
\section{Quantitative Results}
\subsection{Answers to RQ1}
\begin{figure}[htp]
    \centering
    \includegraphics[width=9cm]{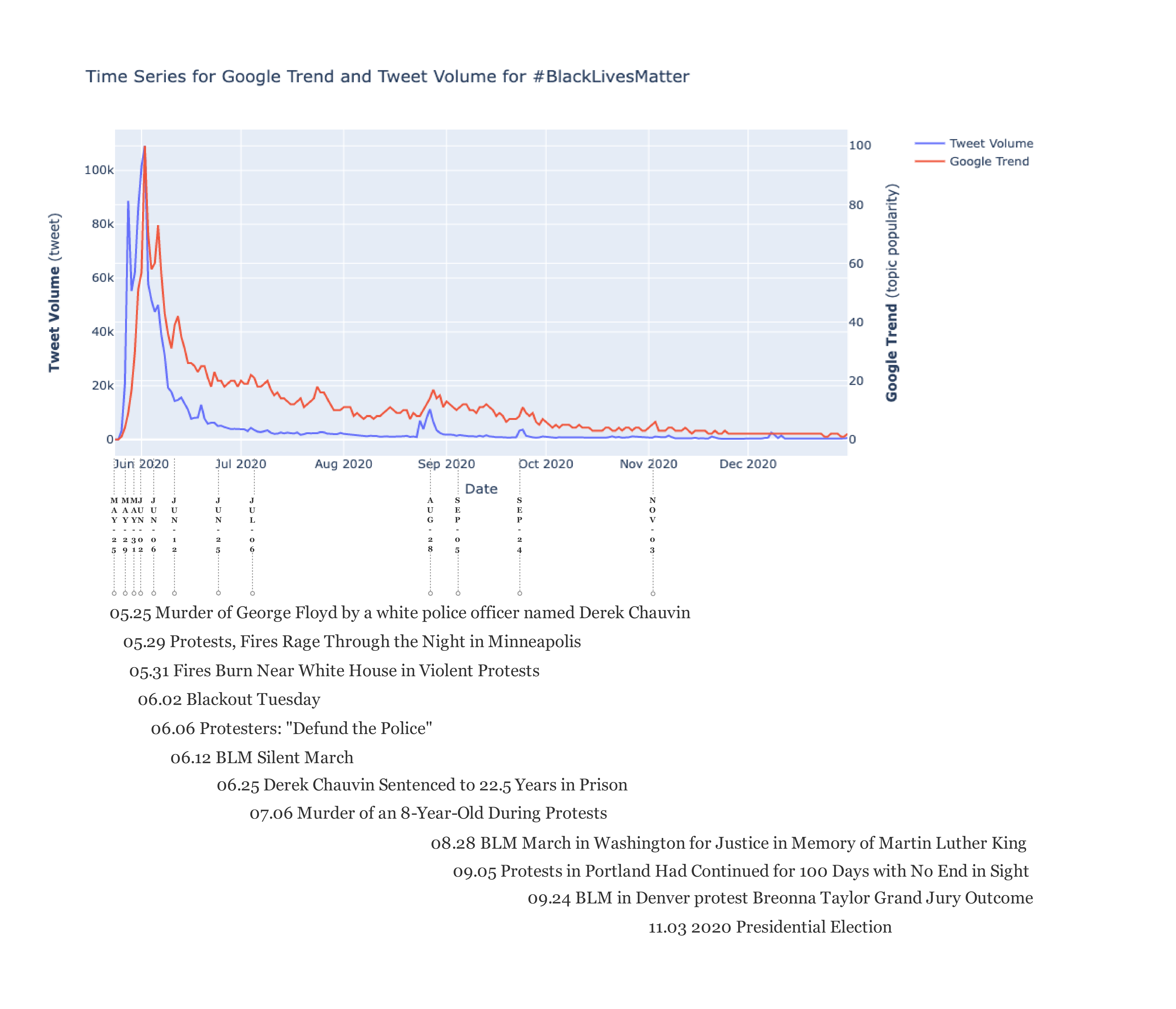}
    \caption {\#Blacklivesmatter:
    Time Series for Google Trend, Tweet Volume, and Events. The red line illustrated the relative interest on Google for the keywords "black lives matter" in the U.S, whereas the blue line represents tweet volumes, the number of tweets per day, and the red line illustrates google trend scores (Timezone: Universal Time Coordinated).
} 
    \label{fig:3}
\end{figure}

To answer this question, we first compared the trends in the popularity of \#blacklivesmatter on Twitter to its popularity on Google. We use  PyTrend\footnote{https://github.com/GeneralMills/pytrends/blob/master/pytrends\\/dailydata.py} to query the Google Trends data during the same time range.
Google Trends data indicate the popularity of various search topics on Google. We compared the trend data on Google and the tweet volume on Twitter and observed a similar and consistent tendency during the timeframe of our dataset (see Figure
~\ref{fig:3}). 
\par
The trends in the two types of data for \#blacklivesmatter seem to be generally consistent: interest skyrockets, peaking in early June, then decreases gradually, bounces back slightly at the end of August, and finally stays at a low level from October on. Moreover, we run a Spearman correlation test on Twitter volume and Google Trends data, and the results showed that they are highly correlated (coefficient= 0.895) with a high significance (P value $<$ 0.001).
\par
\begin{figure}[htp]
    \centering
    \includegraphics[width=9cm]{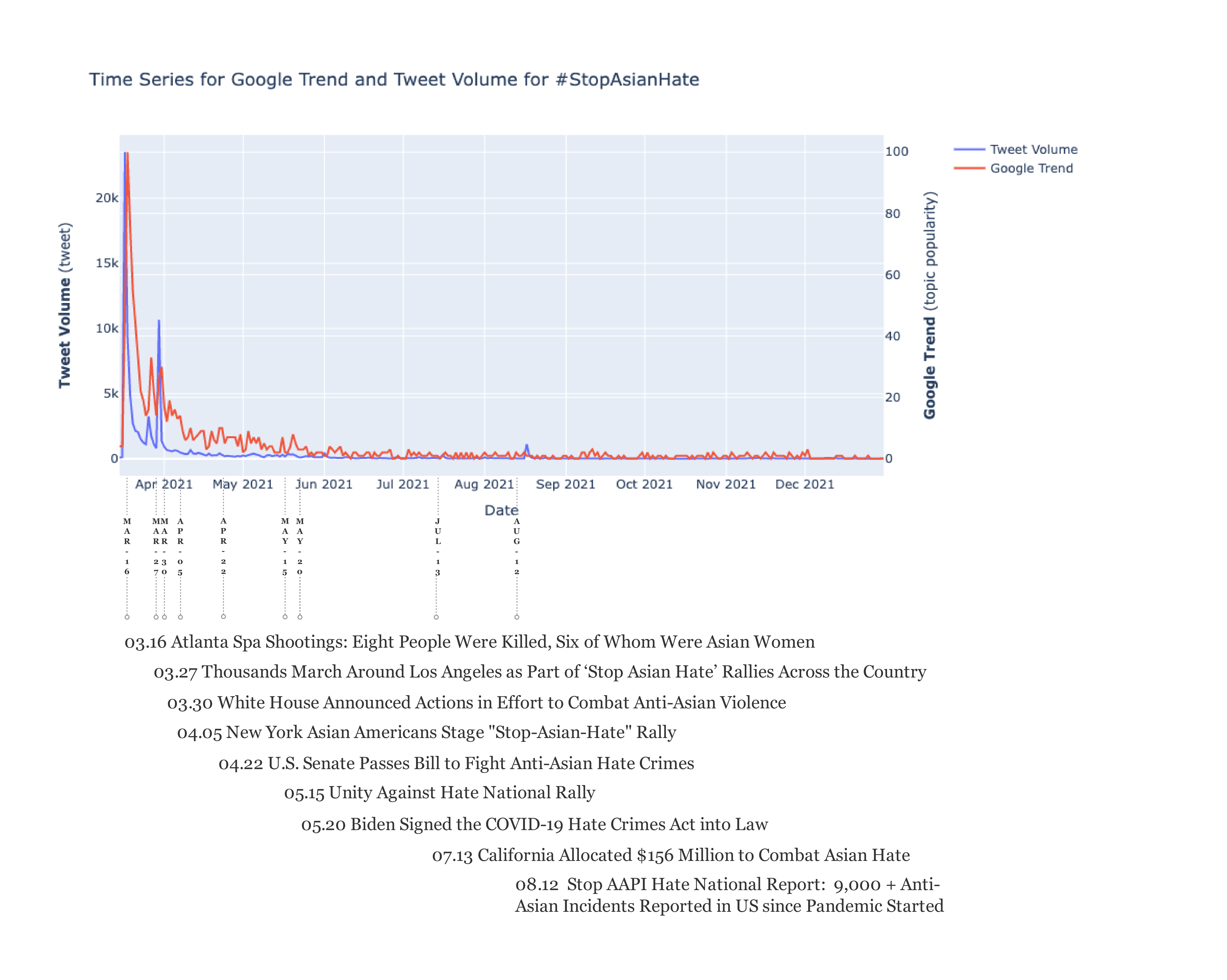}
    \caption {\#Stopasianhate:
    Time Series for Google Trend, Tweet Volume, and Events (spearman correlation coeffient=0.683, p-value$<$0.001)
} 
    \label{fig:4}
\end{figure}

Furthermore, in Figure~\ref{fig:3}, we also observed that each of the peaks in these two online trends coincidentally happen on the same dates there were offline social events, movements, or protests. For example, the first peak in Figure ~\ref{fig:3} is only 1 day after the murder case of George Floyd. Similarly, the second and third peaks occur immediately after the 5.29 and 5.31 protests, respectively, when there were also protests occurred near the White House. We adopted the same approach and analyzed the tweet volume for \#stopasianhate and compared it with the corresponding Google Trend, as shown in Figure ~\ref{fig:4}. Similar to \#blacklivesmatter, the Google Trends data and tweet volume for \#stopasianhate show similar patterns. There were also peaks right after protests, especially for the first two months of the studied period. For instance, the trend/tweet volume reaches the first peak immediately after 3.16 Atlanta spa shooting event. Then, two more peaks emerge a few days before the 3.27 protest and the new actions announced by President Biden on 3.30 to handle anti-Asian activities. 
\par
However, the overall patterns for \#stopasianhate are different than those for \#blacklivesmatter. Unlike for \#blacklivesmatter, the popularity of \#stopasianhate decreases soon after the first several peaks, doing so much faster and staying at a relatively low level of search popularity and tweet volume. However, \#stopasianhate’s Google Trends and tweet volume still show highly significant correlation. This hashtag has a smaller correlation than \#blacklivesmatter data (spearman correlation coefficient=0.679, p-value$<$0.001). 
\par
To conclude, we identified a coincidental correlation between each social movement’s Google Trend and its tweet volume. We also find a close relationship between the dates those protests and social events happened and the dates these two hashtags received the most attention/use. In the appendix, we also include Figure ~\ref{fig:8}, a figure with both the BLM and SAH data. It captures the relative size and time range of tweet conversations.
\subsection{Answers to RQ2}
\subsubsection{Top Words and Network}
We first ranked the overall frequency of words in the tweets to identify top words for both topics (see Figure ~\ref{fig:7} in the Appendix). By ranking the absolute word count for two movements, we find some similarities between the two topics. First, both movements have several associated hashtags. In addition to \#blacklivesmatter and \#stopasianhate, we see other related hashtags, such as \#georgefloyd, \#stopaapihate, \#stopasianhatecrimes, and \#asiansarehuman.
\par
Second, among the top words of both movements, there are calls for participation, such as "please", "support", and "us". There are also words such as “hate”, “racism”, and “violence" that related to racism and prejudice. This finding coincides with \cite{dacon_what_2021}, who concluded that keywords such as "justice for" directly indicate a desire to resolve social issues that affect Black people.
\par
Third, both topics contain phrases indicating a specific group of people. "Black" and "white" in the \#blacklivesmatter movement, "asian", "aapi" and "American" are direct evidence of this. There are also differences in top keywords between the two movements. For instance, "police" is highly ranked in the \#blacklivesmatter data, while “violence” and “community” are highly ranked in the \#stopasianhate data. This finding also coincides with the work by  \cite{dacon_what_2021}, who identified controversial hashtags related to policy- and brutality-related themes in the BLM movement.
\par
After examining the word frequency of tweets, we look into word bigrams to examine how word pairs are connected. Figure ~\ref{fig:10} and Figure ~\ref{fig:11} visualize networks of top 50 co-occurring words in tweets on topic \#Blacklivesmatter and \#Stopasianhate. Each node in the graph represents a word, and each edge represents the connection between two words. The thickness of the edge is a direct indication of word-word co-occurrence frequency. The higher the frequency is, the thicker the edge. Details on the word bigram frequency lists are given in Appendix Table ~\ref{table:4} and Table~\ref{table:5}.
\subsubsection{LDA}
Table ~\ref{tab:3} shows the final topics generated by the LDA model for both the \#blacklivesmatter (Columns 1 and 2) and \#stopasianhate datasets (Columns 3 and 4) with the above mentioned optimal hyperparameters. Out of all final topics (15 for \#blacklivesmatter and 20 \#stopasianhate), we filtered out the topics that had no actual meaning discernable by looking at their top words, which resulted in 12 final meaningful sets of top words for \#blacklivesmatter and 14 for \#stopasianhate (see Table ~\ref{tab:3} for details on topic coverage and keywords). The table shows the topic IDs ranked by topic coverage (Column 1 and Column 3) for both movements, and some represented keywords for each topic. 
\par
Twitter users discussed similar topics in tweet using \#blacklivesmatter and \#stopasianhate, while there were some differences between the results for the two datasets. First, the dominant topic (ID=1) in both movements is movement slogans, accounting for 54.3\% and 44.6\% of BLM and SAH tweets, respectively. Users advocate for social justice while fighting racism, violence, and hate. 
\par
Second, both datasets reference key characters in offline events, including victims, protesters, celebrities, and politicians. Among these, both movements reference victims: 4 of the 12 top BLM topics are the victims George Floyd, Jacob Blake, Tamir Rice, Michael Brown, and Breonna Taylor; 1 of the 14 top SAH topics are the victims Daoyou Feng, Xiaojie Tan, and Soon Chung Park. Interestingly, politicians’ names are topics only within the BLM dataset. Topics 4 and 6, covering 7.7\% of all BLM tweets, include keywords such as Martin Luther King, Joe Biden, Donald Trump, and Kamala Harris. On the other hand, keywords related to celebrities are unique to the SAH dataset. Topics 3, 8, 13, 14, 17, and 18 in SAH include the K-pop group names BTS, Ateez, and GOT7 as well as Emily in Paris star Ashley Park. In addition to specific people, BLM topics also include famous events; the Portland protests, for instance, were a crucial keyword in topic 13 with coverage of 2.3\%. \par
Third, Twitter users expressed similar emotional responses regarding the movements. Some keywords have clear sentiment polarity, either positive or negative, in BLM and SAH tweets. For instance, the keywords "hate" and "oppose" express clear negative polarity; the keywords "support", "love" and "beautiful" express clear positive polarity; and the keywords "mourn", "thanks" and "respect" express attitudes of grief, gratitude, and esteem.
\par
In summary, the main topics identified by LDA modeling depict that users considered Twitter to be a trusted place where they felt comfortable discussing events, attitudes, and challenges regarding racism. The topic and keyword distributions produced by quantitative research inspired our qualitative results in Section 5. 
\section{Qualitative Results}
In this section, we present the qualitative analysis results of the topics being discussed in the selected tweets with the \#blacklivesmatter or \#stopasianhate hashtag. We first introduce the topics that were both popular in both the BLM and the SAH movement on Twitter. Then, we report the topics unique to each movement in Twitter discussions.
\par
We propose a novel approach for analyzing, comparing, and identifying the similarities and differences between the topics discussed on Twitter as part of the BLM and SAH movements. We achieved more robust comparison results by querying tweets under different hashtags independently, whereas a previous study \cite{carney_all_2016} performed qualitative analysis on combined content referenced under multiple hashtags. The LDA model also allowed us to sample tweets based on the distribution of various topics, ensuring comprehensive coverage of the content discussed in the two movements.
\subsection{Common Topics among the Twitter Discussions on the BLM and SAH Social Movements}
\subsubsection{Condemn and fight stop racism.} In both the BLM and SAH movements, approximately 20\% of the sample tweets (113 for the BLM movement and 240 for the SAH movement) condemned racial discrimination and called for social justice. This is because these two movements were triggered by racist hate crimes (i.e., the murder of Black man George Floyd and the shooting death of six Asian women), and the major goal of these two movements is to bring attention to the unequal treatment of and violence against Black and Asian people and stopping racism \cite{taylor_blacklivesmatter_2016, redd_unity_2021}. One tweet in the BLM dataset noted, \textit{“The color of your skin should not determine your value as a human”} (Table~\ref{tab:3}: \#blacklivesmatter - “color”). This post implies that Black people should not be looked down on due to their skin color. It also opposes discriminative behaviors such as committing hate crimes against victims who are chosen according to their social identity \cite{cogan_hate_2002}. People in the SAH movement also posted similar tweets to express their determination and willingness to abolish racism in society: \textit{“We must band [sic: stand] together to continue to fight racism, violence, bigotry and hate twds [towards] ppl [people] of color”} (Table~\ref{tab:3}: \#stopasianhate - “together”, posted by @ngoafulezi). Tweets from both of these movements denounce the unfair treatment of and discrimination against people of color and call for justice.
\subsubsection{Express negative and positive emotions.} People posted tweets to convey both negative (e.g., anger, sorrow, fear) and positive (e.g., pride, gratitude) feelings in relation to the BLM and SAH movements. Although these two movements share the same categories of negative emotions, the dominant emotions within those categories between the two movements are different. When discussing BLM, more users express anger and outrage toward the police, the government, murders, and the overall social situation in the U.S. (refer to Section 4.2 for motivation analysis). For example, one user posted,  \textit{“I am absolutely pissed the f*ck off! They abuse their power every single second!”} (Table~\ref{tab:3}: \#blacklivesmatter - “acab”, posted by @elbowwcheese) to express her dissatisfaction with and accusation against the police for arresting Black people during an offline protest. In contrast, the dominant emotion in SAH discussions was sorrow. Users empathized with victims of hate crimes and with the Asian community. For instance, one user hoped to end the increasingly serious anti-Asian incidents, saying that \textit{“it breaks my heart and makes me sick to think of all the horrible things that have been inflicted upon Asian people.”} (Table~\ref{tab:3}: \#stopasianhate - “people”, posted by @crazyinhwang)
 \par
In addition to negative emotions, users also express positive feelings associated with both the BLM and SAH movements. The positive emotions mainly include the users' pride in the movements' current progress and gratitude toward supporters, such as donors and people who are active in promoting the movements. In the tweet, \textit{“I’m so proud of Houghton. There were more than 1000 people out in the streets today making their voices heard. More were supporting from their cars”} (Table~\ref{tab:3}: \#blacklivesmatter - “today”), the user described citizens actively participating in BLM activities and praised his city for contributing to the BLM movement. As another example, in the SAH movement, one user sent a tweet reading \textit{“Thank you for tuning to our
\#stopasianhate collab! We have raised a goal of USD\$1000 and
I am so happy with everyone’s contribution!”} (Table~\ref{tab:3}: \#stopasianhate - “happy”) to express gratitude to people who have provided financial support to the movement.
\subsubsection{Connections between online and offline activities.}   Our study finds that for both BLM and SAH, the online activity on Twitter and offline events reinforce each other, which answers our first RQ. This result explains the peaks that appeared around the dates of offline activities (e.g., protests, marches) shown in our quantitative results (Figures ~\ref{fig:3} \& ~\ref{fig:4}), and it resonates with \citeauthor{chuang_termite_2012} that offline activities enrich discussions on online social media. Our analysis reveals that on Twitter, people promote future offline events and share helpful resources and tools for organizing events on the online Twitter platform. For example, one tweet read \textit{“Delhi, NY Sat, June 6 at 1 pm. Rally: living, breathing, feeling. Monument to \#georgedloyd and others murdered by racist law enforcement and justice system”} (Table~\ref{tab:3}: \#blacklivesmatter - “pm”); it was posted and retweeted to promote an offline gathering for mourning hate crime victims. Users also post tweets to document and discuss ongoing and past offline events, which are usually shared together with pictures and videos. For instance, a professor recorded his experience participating in an SAH offline rally with a tweet: \textit{“There were around 400 people gathering at Phoenix for \#stopasianhate rally. Both my postdoc Dan and myself are (the) first time to speak out”} (Table~\ref{tab:3}: \#stopasianhate - “speak”, posted by \@joeydai). He also posted four photographs from the event depicting participants, posters, and signs. Similarly, a participant in the BLM movement depicted an organized and successful march that took place in his or her city, saying, \textit{“There was a \#blackLivesMatter rally/march in my city yesterday. Hundreds of people turned out for it and the protesters worked with the police and fire who escorted the march. It was 100\% peaceful!”}  (Table~\ref{tab:3}: \#blacklivesmatter - “peace”). In summary, the online discussions aimed to encourage more people to participate in offline events, and in turn, the offline participants shared their experiences with past events people who were not there.
\par
Our results also imply that the pandemic situation increased the proportion of online activities in the SAH movement more than it did those in the BLM movement. We discovered that there were twice as many tweets related to online events and resource sharing than tweets discussing offline events in the SAH movements but there were half as many tweets documenting offline events as there were documenting online activities for the BLM movement. We assume that this discrepancy is caused by the changing pandemic situation across different years. Since tweets with the \#blacklivesmatter and \#stopasianhate hashtags were crawled from 2020 and 2021, respectively, we infer that there was a stricter lockdown in place when the SAH movement began and thus more online activities emerged.
\subsubsection{ Call for more participation in the movement and building up the community.} Both the BLM and SAH social movements are grown by the two communities spreading awareness, calling for action, and staying united on the Twitter platform. Some users expressed their determination to support people of color and these social movements as well as appealed others to participate in by posting tweets; tweets advocating such activities accounted for approximately 15\% of our collected tweets in both the BLM and SAH datasets. The following tweets are good examples from the SAH and BLM movements, respectively: \textit{“We stand with the Asian American and Pacific Islander community against racism, and it’s time to take action. Please join us in supporting awareness of \#stopAsianHate and learn more (about) how you can help”} (Table~\ref{tab:3}: \#stopasianhate - “against”) and \textit{“I hope we can work together so that everyone can be comfortable in their own skin.”} (Table~\ref{tab:3}: \#blacklivesmatter - “george”). These tweets have a wide target audience since they can make Black people and the Asian community feel supported and comforted while keeping others motivated.
\par
Other users (more than 10\% for both movements) made use of the tag function (@) on Twitter and created a chain by sending a tweet with the \#blacklivesmatter/\#stopasianhate hashtag, tagging a few more users, and telling them to do the same (e.g., “-reply with \#BlackLivesMatter -do the same and tag 6 people: @...”). As a result, knowledge of these two movements spread in a rapid and cost-effective way. As the size of the community increases, the number of people aware of the importance of stopping racism and being called on to participate in the movements also increases. Then, as more people participate in the movement for change, the Black and Asian communities may feel that they are supported and not alone.
\par
In addition to the above topics that both the BLM and SAH movements touched on, each movement had unique topics on Twitter. In Section 4.2, we introduce the qualitative analysis results on tweets discussing the effects of police brutality against Black people, a unique topic that the SAH movement does not share. In Section 4.3, we present distinct topics in the SAH movement, such as anti-Asian hate triggered by the COVID-19 pandemic and support for the movement from celebrities and large business companies.
\subsection{Unique Topic in BLM Twitter Discussions: Conflict with the Police}
Unlike in the SAH movement, anti-police attitudes are an important topic in the BLM movement, where people show their negative emotions regarding police violence. This might be because the killing of Black people by police officers is part of the degradation of Black people in the United States, as the police judge Black people’s behaviors based on their race \cite{tolliver_police_2016}. Additionally, \citeauthor{williamson_black_2018}~\shortcite{williamson_black_2018} find that BLM protests are likely to take place in areas where Black individuals have been slain by police in the past. In contrast, there are much fewer reports linking anti-Asian hate crimes and police brutality, which could explain why this topic did not appear in tweets with the \#stopasianhate hashtag.
\par
Supporters of the BLM movement make three types of accusations against the police and their racist and violent behaviors, i.e., condemnation of the police for committing murders, for interfering in offline protests, and for not doing their duties in general. Within the first type of accusation, people referenced the numerous victims in expressing their discontent toward the police since police brutality is the main cause of murders of Black people. We collected tweets using the \#blacklivesmatter hashtag right after two Black men, George Floyd and Jacob Blake, were murdered and discovered that users posted tweets referencing the murders of several black men, including George Floyd, by the police. For example, one user posted, \textit{“His name was \#georgefloyd and he couldn't breathe. He was a father, and he was murdered by police because of his skin color.”} (Table~\ref{tab:3}: \#blacklivesmatter - “police”). Under the second type of accusation, people were angry with the police for arresting and injuring protesters participating in offline activities as part of the BLM movement, and people believed that the police’s behaviors constituted racism. A tweet reporting that police had interfered with protesters’ behaviors in past offline events said, \textit{“In \#USA police fired tear gas and arrested several \#blackLivesmatter protesters in \#Portland in the early hours of Tuesday morning, as demonstrations against police brutality and racism continued in the city.”} (Table~\ref{tab:3}: \#blacklivesmatter - “police”, posted by @Ruptly). In addition to accusing the police of specific actions, users also indirectly express their disapproval of the police in general because they think that the police fail to fulfill their responsibilities. For example, one user implied that the police weren’t doing their job, saying that \textit{“Police need to be citizen guardians, not militarized enforcers.”} (Table~\ref{tab:3}: \#blacklivesmatter - “police”, posted by @ess\_trainor).
\subsection{Unique Topics in SAH Twitter Discussions}
\subsubsection{Rising anti-Asian hate since the COVID-19 pandemic began.} There were 42 sampled tweets remarking that the number of anti-Asian hate crimes has increased significantly since the outbreak of the COVID-19 pandemic. Such findings are in accordance with the first-reported major COVID-19 outbreak in Wuhan, China, which was believed to be the primary reason for the increasing anti-Asian and anti-Asian American sentiment in the United States \cite{gover_anti-asian_2020}. One sample tweet referred to the 2021 Atlanta spa shootings (citation) and confirmed the growing hatred toward Asians since the pandemic began, noting that \textit{“the horrific shooting in Georgia is just the latest example of the rising tide of violence that Asian Americans have experienced throughout the pandemic”} (Table~\ref{tab:3}: \#stopasianhate - “pandemic”, posted by @NILC). As stated in this tweet, the Atlanta tragedy was confirmed as a racially motivated hate crime specifically targeting Asian women and regarded as just one example of the mounting violence that all Asian communities in the United States are experiencing.
\par
Other tweets point out that racism against Asian people has a long history in the United States, and hostility against Asians has reached a new peak since the inception of the COVID-19 pandemic. For example, one user posted that \textit{“while anti-Asian racism isn't new, it's gotten much worse during the pandemic; addressing its origins while also discussing strategies to move forward is crucial.”}  (Table~\ref{tab:3}: \#stopasianhate - “new”). This tweet confirms that the pandemic has widely fueled anti-Asian sentiment; it further claims that addressing the origin of anti-Asian sentiments is critical in solving certain issues and ending racism against Asian people. In summary, the COVID-19 pandemic is believed to be one of the primary motivations for the increasing hate crimes against Asians, and the relationship between the pandemic and racist acts against Asians has received abundant attention, confirmation, and discussion within the SAH scene.
\subsubsection{Condemn racism triggered by COVID-19 and hope to end the hate.} Our analysis demonstrates that 23 of the sampled tweets condemn individuals who discriminate against Asians and Asian Americans because of the COVID-19 virus, particularly by asserting that Asians should not be blamed for the virus’s spread. One Twitter user stated that \textit{“if y'all followed the precautions then the virus wouldn't (as) be bad as now. Don't blame Asians for it.”} (Table~\ref{tab:3}: \#stopasianhate - “man”). Such voices denouncing racism were often accompanied by messages expressing hope that the growing violence against Asians will quickly cease. Some tweets specifically expressed support for and encouraged people to contact their representatives to vote for the COVID-19 Hate Crimes Act, a bill signed by US President Joe Biden that addresses hate crimes, particularly the increasing number committed against Asian Americans during the pandemic. For example, one user supporting the bill shared updates about the bill, tweeting that \textit{“the House has passed the COVID-19 Hate Crimes Act with an overwhelming bipartisan majority and it's now headed to President Biden's desk. I was proud to co-sponsor and vote for this important bill to support our AAPI community and help \#StopAsianHate.”} (Table~\ref{tab:3}: \#stopasianhate - “bill”). The tweet not only gave an update about the status of the relevant legislation but also documented the user’s personal participation (e.g., cosponsoring and voting for the bill) in the movement. Whether by expressing support or proposing tangible actions, these tweets showed a strong tendency to advocate for relevant policies and support the Asian community in the context of the pandemic.
\subsubsection{Support for the AAPI community from Asian celebrities and large corporations.} We observed that several Asian celebrities in the entertainment industry, such as South Korean boy bands (e.g., BTS, Ateez), South Korean American actors and producers (e.g., Daniel Dae Kim) and Chinese American actors, directors and producers (Daniel Wu) as well as large businesses, such as Tom Ford, Comcast NBC, IBM, and Burger King, expressed their support for the AAPI community by posting on Twitter.
\par
Many Asian celebrities in the entertainment industry tweeted their support for the Asian community and help raise awareness about the challenges that Asians face today. For example, the South Korean boy band BTS tweeted the following on their official account: \textit{“We stand against racial discrimination. We condemn violence. You, I and we all have the right to be respected. We will stand together”}. Due to celebrities’ fame, their posts receive a great amount of feedback from their fans and followers, which soon leads to the viral circulation of their content as fans and followers extensively retweet the original tweets. According to Twitter, BTS’s \#stopasianhate tweet was the most retweeted tweet and the second most liked tweet in 2021. We found 47 sampled tweets that responded to the celebrities’ tweets, saying that they would stand together with the celebrities and Asian people suffering from racial discrimination. One sample tweet noted that \textit{“I'm both comforted by BTS's words and also incredibly livid at everything both past and present. Either way, it's BTS appreciation hours and it's long overdue for anti-Asian rhetoric and harassment to end”} (Table~\ref{tab:3}: \#stopasianhate - “end”). The celebrities’ reassuring words elicited a flood of positive responses and active engagement from their fans, extending the SAH movement’s social influence.
\par
Similarly, several large corporations also publicly expressed their support on Twitter for the SAH movement and declared their determination to stand with the Asian community. For example, Tom Ford’s official Twitter account posted the message \textit{“We are devastated by the recent acts of violence and bigotry in the US. We stand united with the Asian community in strongly condemning all forms of hate. TOM FORD remains committed to an inclusive future that celebrates and honors all of our differences”} (~\ref{tab:3}: \#Stopasianhate - “condemn”, posted by @TOMFORD) to express support for the SAH movement. Asian celebrities and large corporations, as some of the most influential voices on social media, have readily accepted their social responsibilities and encouraged their audiences to support and participate in the SAH movement online.
\section{Discussion and Conclusions}
We identified three major findings regarding the similarities and differences between the BLM movement and the SAH movement by evaluating and merging the quantitative and qualitative results. 
\par
First, both quantitative and qualitative findings show that online social movements on Twitter and offline events in both movements reinforce one another. The time series results (Figures ~\ref{fig:3} \& ~\ref{fig:4}) demonstrate that tweet volume and Google searches peak around the time of significant offline activities in both movements, indicating that BLM and SAH follow the same pattern of coinciding online and offline peaks that many other online social movements do~\cite{tufekci_twitter_2017,papacharissi_affective_2014}. Furthermore, our qualitative analysis reveals that offline events and online activities have a mutually constructive relationship. On the one hand, people use online social platforms to promote forthcoming offline events. On the other hand, information about past or ongoing offline events is extensively shared on online platforms, increasing the overall online engagement.
\par
Second, we anticipated that the disparity between the topics in the two movements stems from these movements' differing origins. Our qualitative analysis indicates that whereas many voices in the BLM movement focus on police brutality against Black people, dialogues in the SAH movement seek to address the rising racism against Asians since COVID-19 began. We hypothesized that the discrepancy was due to the two movements' varying origins, with the majority of people unjustly killed by the police being Black~\cite{alang_police_2017} and the increasing anti-Asian sentiment deriving from the first case being reported in China ~\cite{gover_anti-asian_2020}.
\par
Finally, we observed that discussions in the SAH movement cover a broader range of topics than the BLM movement. The LDA model with the highest coherence scores yielded 15 topics for the BLM movement and 20 topics for the SAH movement in the quantitative analysis, indicating that the content of the SAH movement conversation is richer than that of the BLM movement. This finding was also validated by the qualitative analysis, in which we identified 42 topic categories for the BLM movement and 87 for the SAH movement. Further investigation is required to uncover the underlying reasons for the difference in topics between the BLM and SAH movements.
\par
Our research focused on the interplay of AI, ethics, and society. First, a critical stream of cooperative AI research seeks to design algorithms to steer social media so as to promote healthy online communities~\cite{dacon_what_2021}. Our analysis shows that the microblogging on Twitter related to BLM and SAH has been an essential part of both these social movements to support minorities, and this activity is supported by the hashtag algorithms on Twitter~\cite{li_twitter_2011}. However, most tweets do not contain tags that impedes the potential formation of connections among online social movement communities. Future research can design automated content-based hashtag recommendations for tweets to further connect online communities and support social movements on Twitter~\cite{godin_using_2013}. Second, fairness and justice are essential facets of AI and ethics research~\cite{whittlestone_role_2019}. Social movements have a long relationship with social justice and fairness~\cite{thompson_social_2002,tyler_social_1995}, which concurs with our results. For example, our LDA analysis shows that one common topic between the BLM and SAH movements is racism, and combatting racism is an essential part of fairness and justice~\cite{overby_justice_2004}. Future work could seek to identify Twitter algorithms that promote diversity and inclusion by reducing racism and discrimination. Finally, our society is undergoing a transition to a cyber–physical–social system (CPSS), in which the virtual, the physical, and the social are evolving into one social space~\cite{xiong_cyber-physical-social_2015}. Our topic analysis documents the Twitter platform as a miniature CPSS. Future research could explore how to design social media platforms that serve as a CPSS that is secure, fair, and efficient~\cite{tan_blockchain-based_2020}.
\par
We recognize a few limitations in our study. First, the Twitter data we queried cover only a few months. Specifically, our data query range is 8 months for BLM and 10 months for SAH, as we used the day after George Floyd’s death and the day after the Atlanta spa shooting as the beginning query dates and the end of 2020 and 2021 as the ending dates, respectively. As a result, these query ranges include the peaks in engagement for the aforementioned two events according to the Google Trends data but fail to cover the whole storyline of either movements. Second, we set a threshold of three for the number of likes to filter tweets, aiming to decrease the computational costs and select tweets that are relevant to the movements, which were randomly selected since prior studies did not indicate a rational method for deciding this threshold. It is possible that tweets with fewer than three likes also contain valuable information. Third, not all our tweets are in English even though we set the snscrape API to query English tweets. When performing the qualitative analysis, we also found Spanish tweets in our dataset. This influenced our qualitative work since we had to translate Spanish into English to understand the tweets’ content before the analysis. Fourth, our study does not include ethnographical work. We conducted quantitative and qualitative research based on the content of the queried tweets. However, we do not take the Twitter users’ demographical information (e.g., race, social class, gender, age, income) into consideration, nor did we reach out to them for interviews. A better understanding of the tweet content could be achieved through close contact with the people who sent these tweets.
\par
Our work makes two major contributions. First, our analysis of the connections between online and offline movements and Twitter topic content sheds light on the importance of social media, which links individuals who share common concerns. Second, we identified the main themes in the BLM and SAH online social movements that cover the major topics of discussion on Twitter through quantitative (i.e., the LDA model, network) and qualitative (i.e., open coding) methods. These themes could help future researchers understand the commonalities and differences between the content of Twitter discussions around these two movements.
\bibliographystyle{ACM-Reference-Format} 
\bibliography{bib_final.bib}

\section{Appendix}

\begin{table}[htp]
\setlength{\tabcolsep}{1.5mm}{
\begin{tabular}{ |c | c | c | c | } 
\hline
   \thead{Alpha} &\thead{Beta} & \thead{\makecell{\#Blacklivesmatter\\ Coherence\\ (\#Topics)}} & \thead{\makecell{\#Stopasianhate\\ Coherence\\(\#Topics)}} \\ 
\hline
0.01 & 0.91 & 0.451 (\#15)  & 0.408(\#20) \\ 
0.31 & 0.91 & 0.512 (\#15)  & 0.483\#20)\\ 
\hline
symmetric & 0.91 & 0.505 (\#15) & 0.4\#31(\#20)\\ 
\hline
0.61 & 0.91 & 0.481 (\#15) & 0.405 (\#20)\\ 
\hline
0.31 & 0.61 & 0.504 (\#15) & 0.468(\#20) \\ 
\hline
0.01 & 0.61 & 0.479 (\#15) & 0.386 (\#20)\\ 
\hline
symmetric & 0.61 & 0.494 (\#15) & 0.464 (\#20)\\ 
\hline
\end{tabular}
\caption{Model Parameters Tuning}
\label{tab:2}}
\end{table}

\begin{figure}[htp]
    \centering
    \includegraphics[width=8cm]{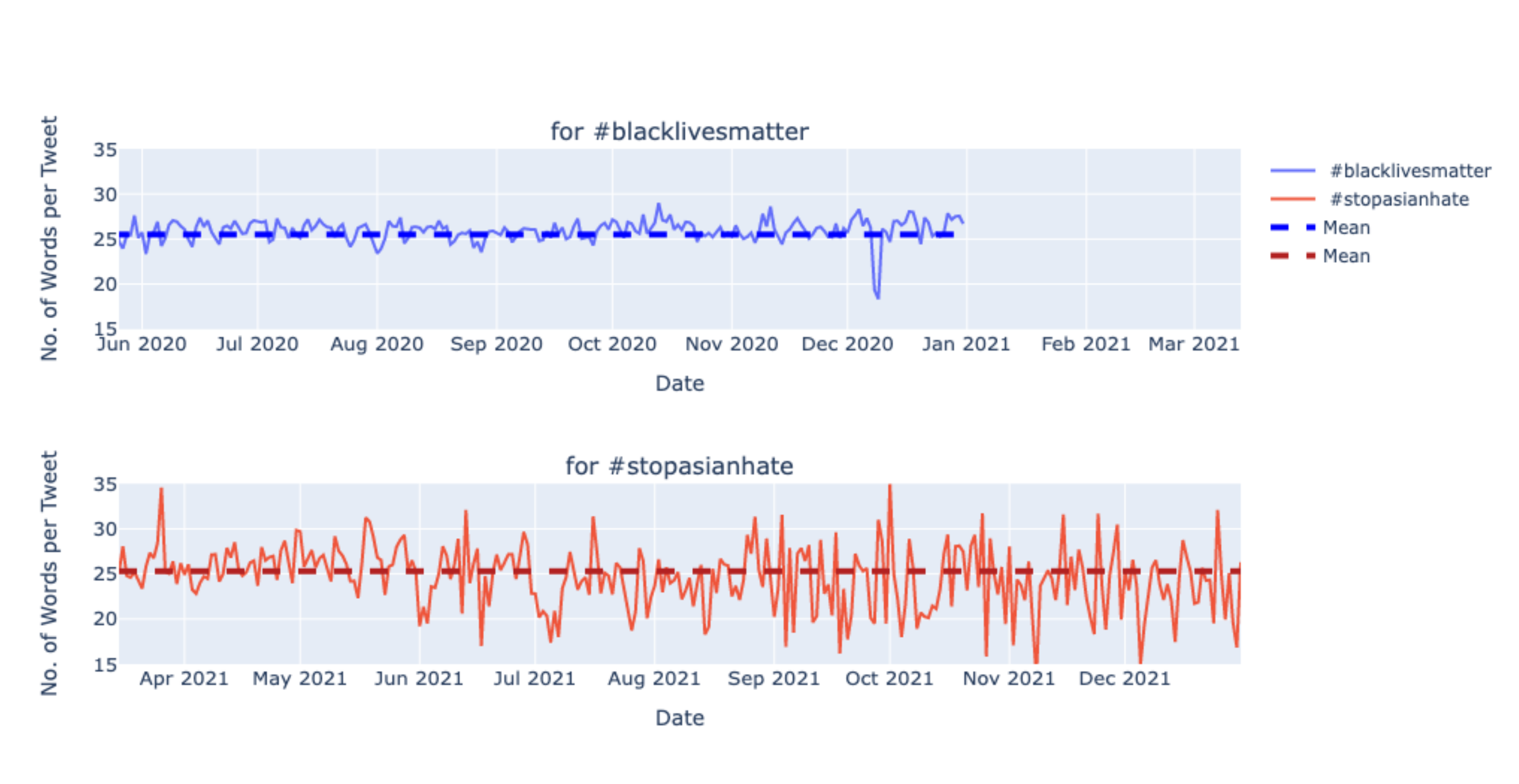}
    \caption {Time Series for Average Tweet Length (daily)}
    \label{fig:1}
\end{figure}

Figure ~\ref{fig:1} shows the average word lengths of \#blacklivesmatter (top) and \#stopasianhate (bottom) tweets over the studied period. For Dataset B, the tweets have an average of 25.3 words (standard deviation=15.0), and the daily averages vary between 1 and 243 words. Similarly, in Dataset S, the average length is 24.9 words (standard deviation=15.2), and the daily average varies between 1 and 144 words. The top 20 words for all \#blacklivesmatter tweets are \#blacklivesmatter (1266949), black (197853), people (157026), lives (78711), police (76502), us (62850), racism (62034), support (60291), matter (58572), white (57644), please (57032), today (55573), i’m (54411), que (54037), see (53572), \#georgefloyd (51808), protest (60641), justice (49691), like (447970), and know (47461). The top 20 words of all \#stopasianhate tweets are: \#stopasianhate (96977), asian (20523), \#stopaapihate (16458), hate (14018), community (9282), racism (9235), violence (8566), stand (8009), people (7847), us (6717), stop (6338), \#stopasianhatecrimes (6245), que (6098), asians (5278), anti-asian (5050), aapi (5004), \#asiansarehuman (4926), please (4833), support (4815), and American (4640).

\begin{figure}[htp]
    \centering
    \includegraphics[width=7cm]{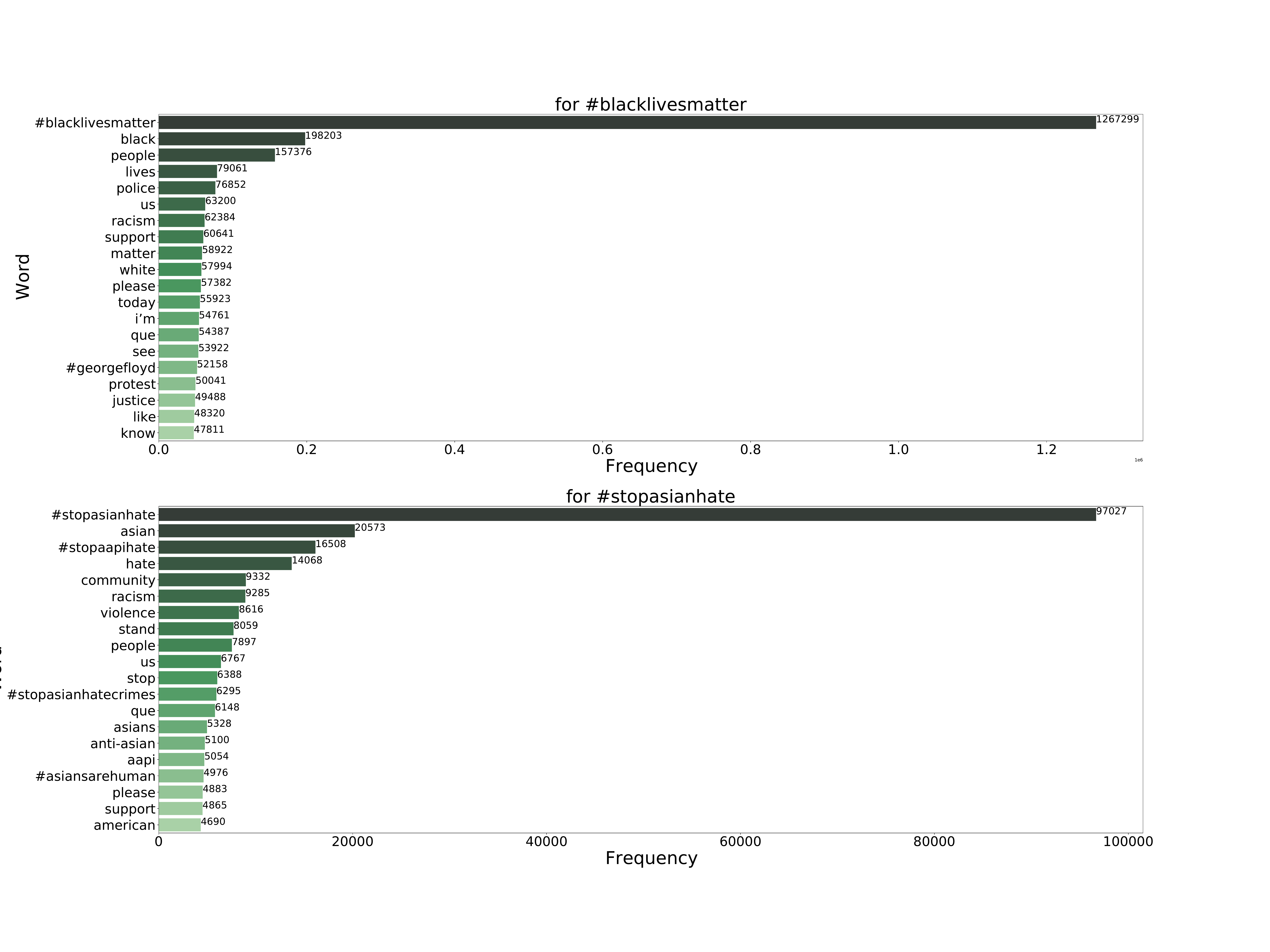}
    \caption {Overall Word Frequency Ranking}
    \label{fig:7}
\end{figure}

\begin{figure}[htp]
    \centering
    \includegraphics[width=8cm]{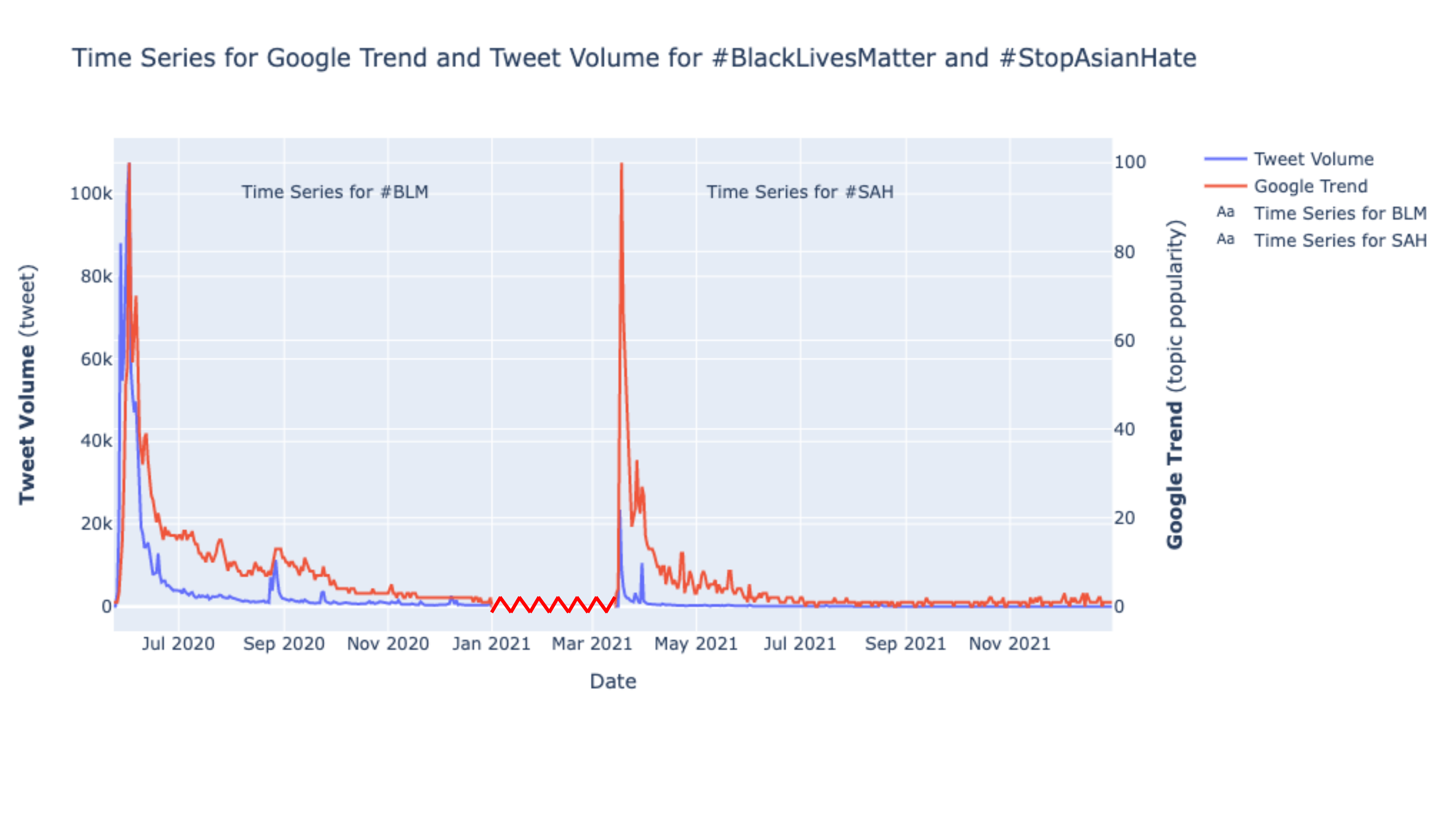}
    \caption {Time Series of Google Trend and Tweet Volume for \#BlackLivesMatter and \#StopAsianHate}
    \label{fig:8}
\end{figure}

\begin{figure}[htp]
    \centering
    \includegraphics[width=8cm]{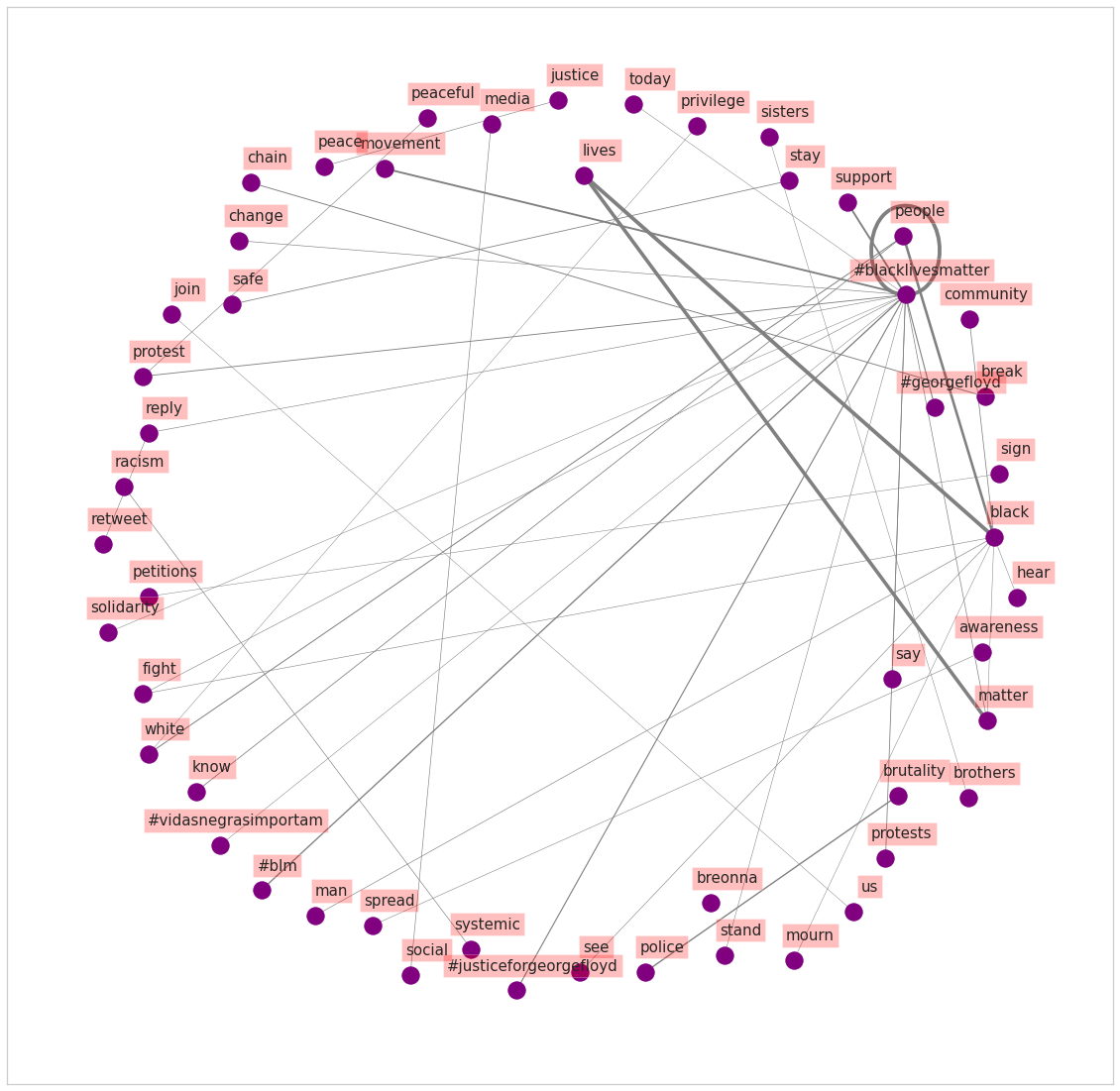}
    \caption {\#Blacklivesmatter:Networks of top 50 co-occurring words in tweets
} 
    \label{fig:10}
\end{figure}

\begin{figure}[htp]
    \centering
    \includegraphics[width=8cm]{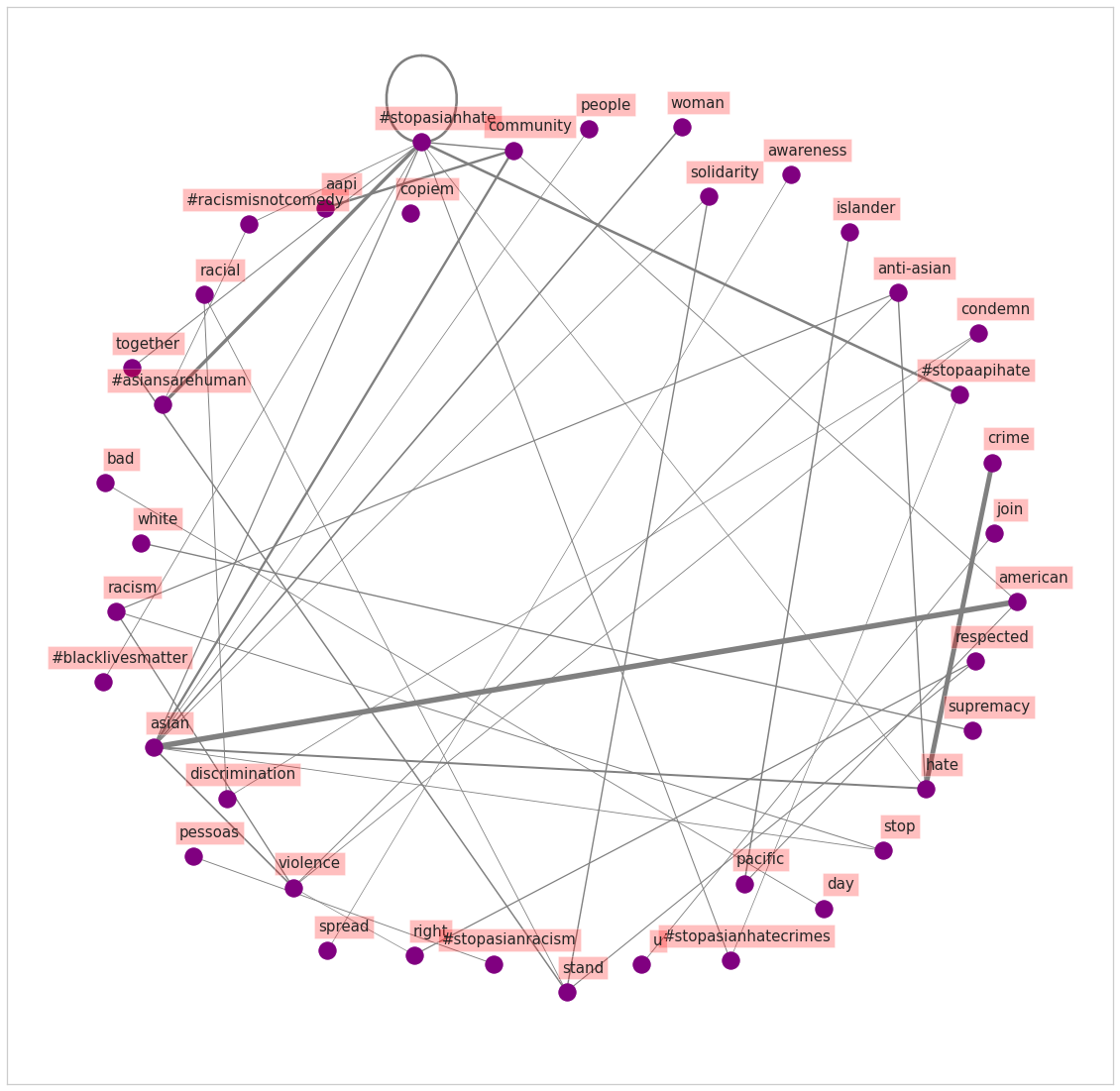}
    \caption {\#Stopasianhate: Networks of top 50 co-occurring words in tweets
} 
    \label{fig:11}
\end{figure}

Figure ~\ref{fig:10} shows the network of top 50 co-occurring words in \#blacklivesmatter twitter discussion. In addition to related hashtags such as \#justiceforgeorgefloyd and \#blm, the term “police brutality” has a word co-occurrence of 13649 among all the BLM tweets.In terms of degree centrality, "black" has 9 edges; "matter" and "people" have 3 edges; "lives", "white", "protest", "reply" and "fight" have 2 edges. These words related to protest slogans have a relatively high degree centrality and positive signals in advocating for the BLM movement.
\par
Figure ~\ref{fig:11} shows the network of top 50 co-occurring words in \#stopasianhate twitter discussion. The SAH dataset has the following top-ranking hashtags: \#stopaapihate, \#asiansarehuman, \#stopasianhatecrimes, \#racismisnotcomedy, \#stopasianhatecrimes". In addition, asian american, asian community and asian woman are co-occurring words ranked in the top 10 word bigram. In terms of degree centrality, "asian" has 9 edges; "violence" has 5 edges; "hate", "community" and "stand" have 4 edges; "american", "anti-asian" and "racism" have 3 edges; "\#stopaapihate", "\#asiansarehuman", "\#stopasianhatecrimes", "pacific", "together", "solidarity", "\#racismisnotcomedy", "right", "respected", "stop", "racial", "discrimination" and "condemn" have 2 edges. Similar to the BLM words, the SAH words with a high degree of centrality are linked to support for the movement and addresses racial issues. In addition, multiple central nodes in the SAH dataset are words that directly express negative attitudes toward the incidents.

\begin{table}[htp]
\setlength{\tabcolsep}{1mm}{
\begin{tabular}{ | c |c| c| c | } 
\hline
 \thead{ID} & \thead{Frequency} &\thead{Type} & \thead{Description} \\ 
\hline
Date & daily & \makecell{ YYYY-MM-DD} & \makecell{\#Blacklivesmatter: from \\May 24, 2020,\\ to Dec 31, 2020
\\\#Stopasianhate: from\\ March 15, 2021,\\ to Dec 31, 2021}
 \\ 
\hline
Tweets & tweet & String & \makecell{Tweets that contain\\ the hashtag\\ \#blacklivesmatter \\ the number of likes $\geq$ 3}\\ 
\hline
\end{tabular}
\caption{Data Dictionary}
\label{table:1}}
\end{table}

\begin{table}[htp]
\begin{tabular}{|l|l|l|}
\hline
 & bigram & count \\ \hline
0 & ('black', 'lives') & 38087 \\ \hline
1 & ('lives', 'matter') & 35761 \\ \hline
2 & ('black', 'people') & 25622 \\ \hline
3 & ('george', 'floyd') & 25255 \\ \hline
4 & ('\#blacklivesmatter', 'movement') & 18917 \\ \hline
5 & ('support', '\#blacklivesmatter') & 18590 \\ \hline
6 & ('\#blacklivesmatter', '\#justiceforgeorgefloyd') & 16504 \\ \hline
7 & ('\#blacklivesmatter', '\#blacklivesmatter') & 15742 \\ \hline
8 & ('police', 'brutality') & 13649 \\ \hline
9 & ('\#blacklivesmatter', '\#blm') & 13279 \\ \hline
10 & ('\#blm', '\#blacklivesmatter') & 13219 \\ \hline
11 & ('\#blacklivesmatter', '\#georgefloyd') & 12886 \\ \hline
12 & ('\#justiceforgeorgefloyd', '\#blacklivesmatter') & 11279 \\ \hline
13 & ('\#georgefloyd', '\#blacklivesmatter') & 10998 \\ \hline
14 & ('white', 'people') & 9851 \\ \hline
15 & ('break', 'chain') & 9370 \\ \hline
16 & ('\#blacklivesmatter', 'protest') & 8809 \\ \hline
17 & ('\#blacklivesmatter', 'protests') & 8041 \\ \hline
18 & ('stay', 'safe') & 7975 \\ \hline
19 & ('people', 'know') & 7897 \\ \hline
20 & ('breonna', 'taylor') & 7526 \\ \hline
21 & ('black', 'community') & 7407 \\ \hline
22 & ('matter', '\#blacklivesmatter') & 6868 \\ \hline
23 & ('systemic', 'racism') & 6460 \\ \hline
24 & ('black', 'see') & 6408 \\ \hline
25 & ('say', '\#blacklivesmatter') & 6354 \\ \hline
26 & ('black', 'man') & 6281 \\ \hline
27 & ('social', 'media') & 6197 \\ \hline
28 & ('change', '\#blacklivesmatter') & 6107 \\ \hline
29 & ('justice', 'peace') & 6035 \\ \hline
30 & ('peaceful', 'protest') & 5864 \\ \hline
31 & ('reply', '\#blacklivesmatter') & 5861 \\ \hline
32 & ('black', 'hear') & 5854 \\ \hline
33 & ('black', 'mourn') & 5797 \\ \hline
34 & ('retweet', 'reply') & 5759 \\ \hline
35 & ('fight', '\#blacklivesmatter') & 5716 \\ \hline
36 & ('see', 'black') & 5715 \\ \hline
37 & ('black', 'fight') & 5492 \\ \hline
38 & ('spread', 'awareness') & 5275 \\ \hline
39 & ('stand', '\#blacklivesmatter') & 5230 \\ \hline
40 & ('hear', 'black') & 5069 \\ \hline
41 & ('\#blacklivesmatter', '\#vidasnegrasimportam') & 5017 \\ \hline
42 & ('matter', 'black') & 4922 \\ \hline
43 & ('solidarity', '\#blacklivesmatter') & 4855 \\ \hline
44 & ('sign', 'petitions') & 4847 \\ \hline
45 & ('today', '\#blacklivesmatter') & 4809 \\ \hline
46 & ('brothers', 'sisters') & 4806 \\ \hline
47 & ('mourn', 'black') & 4712 \\ \hline
48 & ('white', 'privilege') & 4562 \\ \hline
49 & ('join', 'us') & 4515 \\ \hline
\end{tabular}
\caption{Word Bigram Ranking for \#Blacklivesmatter}
\label{table:4}
\end{table}

\begin{table}[htp]
\begin{tabular}{|l|l|l|}
\hline
 & bigram & count \\ \hline
0 & ('\#stopasianhate', '\#stopaapihate') & 9538 \\ \hline
1 & ('asian', 'american') & 5519 \\ \hline
2 & ('hate', 'crime') & 4858 \\ \hline
3 & ('\#stopasianhate', '\#asiansarehuman') & 3429 \\ \hline
4 & ('\#stopasianhate', '\#stopasianhatecrimes') & 2768 \\ \hline
5 & ('\#stopaapihate', '\#stopasianhate') & 2556 \\ \hline
6 & ('aapi', 'community') & 2408 \\ \hline
7 & ('asian', 'community') & 2323 \\ \hline
8 & ('asian', 'hate') & 1998 \\ \hline
9 & ('asian', 'woman') & 1645 \\ \hline
10 & ('violence', 'asian') & 1644 \\ \hline
11 & ('pacific', 'islander') & 1511 \\ \hline
12 & ('stand', 'together') & 1480 \\ \hline
13 & ('community', '\#stopasianhate') & 1429 \\ \hline
14 & ('anti-asian', 'hate') & 1388 \\ \hline
15 & ('stand', 'solidarity') & 1378 \\ \hline
16 & ('white', 'supremacy') & 1360 \\ \hline
17 & ('racism', 'violence') & 1313 \\ \hline
18 & ('\#racismisnotcomedy', '\#stopasianhate') & 1275 \\ \hline
19 & ('right', 'respected') & 1251 \\ \hline
20 & ('\#stopaapihate', '\#stopasianhatecrimes') & 1237 \\ \hline
21 & ('asian', '\#stopasianhate') & 1206 \\ \hline
22 & ('anti-asian', 'racism') & 1189 \\ \hline
23 & ('\#stopasianhate', '\#stopasianhate') & 1104 \\ \hline
24 & ('respected', 'stand') & 1077 \\ \hline
25 & ('together', '\#stopasianhate') & 1043 \\ \hline
26 & ('\#stopasianhatecrimes', '\#stopasianhate') & 1022 \\ \hline
27 & ('american', 'pacific') & 1021 \\ \hline
28 & ('stop', 'racism') & 1001 \\ \hline
29 & ('racial', 'discrimination') & 961 \\ \hline
30 & ('anti-asian', 'violence') & 958 \\ \hline
31 & ('tag', 'pfv') & 935 \\ \hline
32 & ('copiem', 'tag') & 924 \\ \hline
33 & ('pessoas', '\#stopasianracism') & 917 \\ \hline
34 & ('american', 'community') & 904 \\ \hline
35 & ('join', 'u') & 871 \\ \hline
36 & ('bad', 'day') & 864 \\ \hline
37 & ('condemn', 'violence') & 845 \\ \hline
38 & ('stop', 'asian') & 839 \\ \hline
39 & ('\#stopasianhate', '\#racismisnotcomedy') & 831 \\ \hline
40 & ('stand', 'racial') & 825 \\ \hline
41 & ('\#blacklivesmatter', '\#stopasianhate') & 821 \\ \hline
42 & ('solidarity', 'asian') & 820 \\ \hline
43 & ('\#asiansarehuman', '\#racismisnotcomedy') & 797 \\ \hline
44 & ('discrimination', 'condemn') & 768 \\ \hline
45 & ('asian', 'people') & 746 \\ \hline
46 & ('hate', '\#stopasianhate') & 715 \\ \hline
47 & ('\#stopasianhatecrimes', '\#stopaapihate') & 711 \\ \hline
48 & ('violence', 'right') & 682 \\ \hline
49 & ('spread', 'awareness') & 681 \\ \hline
\end{tabular}
\caption{Word Bigram Ranking for \#Stopasianhate}
\label{table:5}
\end{table}

\newpage
\begin{table}[htp]
\scalebox{0.75}{
\begin{tabular}{|ll|ll|}
\hline
\multicolumn{2}{|l|}{\#Blacklivesmatter} & \multicolumn{2}{l|}{\#Stopasianhate} \\ \hline
\multicolumn{1}{|l|}{\begin{tabular}[c]{@{}l@{}}Topic ID\\  (coverage)\end{tabular}} & Key words & \multicolumn{1}{l|}{\begin{tabular}[c]{@{}l@{}}Topic ID \\ (coverage)\end{tabular}} & Top words \\ \hline
\multicolumn{1}{|l|}{\begin{tabular}[c]{@{}l@{}}1\\ (54.3\%)\end{tabular}} & \begin{tabular}[c]{@{}l@{}}black, people, police, \\ blm, lives, us, racism, \\ today, matter, justice\end{tabular} & \multicolumn{1}{l|}{1(44.6\%)} & \begin{tabular}[c]{@{}l@{}}stopasianhate, asian, \\ stopaapihate,hate, aapi, \\ community, racism, \\ stand, anti, violence, covid\end{tabular} \\ \hline
\multicolumn{1}{|l|}{\begin{tabular}[c]{@{}l@{}}2\\ (5.4\%)\end{tabular}} & \begin{tabular}[c]{@{}l@{}}racismo, movimento, \\ quando, negra\end{tabular} & \multicolumn{1}{l|}{2(11.8\%)} & \begin{tabular}[c]{@{}l@{}}stopasianhate, people, asian, \\ hate, black, us, asians, stop, \\ racist, please\end{tabular} \\ \hline
\multicolumn{1}{|l|}{3(4.9\%)} & \begin{tabular}[c]{@{}l@{}}jacob, para, racismo, \\ contra, policia, \\ movimiento\end{tabular} & \multicolumn{1}{l|}{3(9.6\%)} & \begin{tabular}[c]{@{}l@{}}stopasianhate, stopaapihate,\\  como, todos, bts, \\ racismo, bts\_twt\end{tabular} \\ \hline
\multicolumn{1}{|l|}{4(4.1\%)} & \begin{tabular}[c]{@{}l@{}}breonna, taylor, racism,\\  martin, michael, france, \\ luther\end{tabular} & \multicolumn{1}{l|}{6(2.4\%)} & \begin{tabular}[c]{@{}l@{}}stopasianhate, asian, month, \\ aapi, heritage, community, \\ students, americans, \\ atlanta, support\end{tabular} \\ \hline
\multicolumn{1}{|l|}{5(3.7\%)} & \begin{tabular}[c]{@{}l@{}}retweet, petition, tag, \\ reply, mourn\end{tabular} & \multicolumn{1}{l|}{8(2\%)} & \begin{tabular}[c]{@{}l@{}}stopasianhate, paul, chung,\\ xiaojie, tan, che, daoyou, \\ ashley, park, andre\end{tabular} \\ \hline
\multicolumn{1}{|l|}{6 (3.6\%)} & \begin{tabular}[c]{@{}l@{}}bidenharris, joebiden,\\ realdonaldtrump,\\ charged, breonnataylor, \\ georgefloyd, nyc, \\ kamalaharris\end{tabular} & \multicolumn{1}{l|}{9(2\%)} & \begin{tabular}[c]{@{}l@{}}stopasianhate, racism, stop,\\ respect, asians, beautiful, \\ francisco, violence, people\end{tabular} \\ \hline
\multicolumn{1}{|l|}{7(3.5\%)} & \begin{tabular}[c]{@{}l@{}}black, lives, matter, \\ hear, women, trans, \\ skin, brothers, sisters,\\  brown\end{tabular} & \multicolumn{1}{l|}{12(1.8\%)} & \begin{tabular}[c]{@{}l@{}}stopasianhate, asian, brothers,\\  sisters, white, stopaapihate,\\  racism, birthday, supremacy\end{tabular} \\ \hline
\multicolumn{1}{|l|}{8(3\%)} & \begin{tabular}[c]{@{}l@{}}break, chaim, sportify\\ blackexcellence, papua\end{tabular} & \multicolumn{1}{l|}{13(1.7\%)} & \begin{tabular}[c]{@{}l@{}}stopasianhate, please, \\ asiansarehuman, racism,\\  bts\_twt, retweet, \\ time, take, violence\end{tabular} \\ \hline
\multicolumn{1}{|l|}{10 (2.9\%)} & \begin{tabular}[c]{@{}l@{}}blake, repsect, jacobblake, \\ wethenorth,\end{tabular} & \multicolumn{1}{l|}{14(1.6\%)} & \begin{tabular}[c]{@{}l@{}}stopasianhate, ateez, \\ asiansarehuman, \\ racism, bts\_twt\end{tabular} \\ \hline
\multicolumn{1}{|l|}{11(2.6\%)} & \begin{tabular}[c]{@{}l@{}}pledge, allyship, oppose, \\ hate, thanks, active,\\ learning, quote, \\ listening, smith\end{tabular} & \multicolumn{1}{l|}{16(1.5\%)} & \begin{tabular}[c]{@{}l@{}}stopasianhate, asian, \\ kpixtv, love, hate, \\ chicago, community, \\ stopaapihate, people\end{tabular} \\ \hline
\multicolumn{1}{|l|}{12(2.5\%)} & \begin{tabular}[c]{@{}l@{}}notoracism, california,\\  memorial, saynotoracism, \\ naomiosaka, tamirrice\end{tabular} & \multicolumn{1}{l|}{17(1.5\%)} & \begin{tabular}[c]{@{}l@{}}stopasianhate, lin, fans,\\ respeitados(respect), \\  asiansarehuman, yoogi, got\end{tabular} \\ \hline
\multicolumn{1}{|l|}{13(2.3\%)} & \begin{tabular}[c]{@{}l@{}}portland, portlandprotest, \\ pdxprotests, oregon, \\ pdxproetst, pdx, \\ defendpdx, \\ portlandoregon, \\ portlandstrong,\\ wallofmoms\end{tabular} & \multicolumn{1}{l|}{18(1.4\%)} & \begin{tabular}[c]{@{}l@{}}stopasianhate, film, \\ chesaboudin, studies, \\ asian, nba, hate, clubhouse\end{tabular} \\ \hline
\multicolumn{2}{|l|}{} & \multicolumn{1}{l|}{19(1.4\%)} & \begin{tabular}[c]{@{}l@{}}stopasianhate, stopaapihate, \\ asian, community, match, \\ asiansarehuman, endracism, \\ stand, violence\end{tabular} \\ \hline
\multicolumn{2}{|l|}{} & \multicolumn{1}{l|}{20(1.4\%)} & \begin{tabular}[c]{@{}l@{}}stopasianhate, asian, racism, \\ violence, missuniverse, \\ asiansarehuman, white, \\ woman, toravines\end{tabular} \\ \hline
\end{tabular}}
\caption{Sentiment Analysis: Calculated Variables}
\label{tab:3}
\end{table}

\end{document}